\documentclass[%
aip,
amsmath,amssymb,
reprint,%
]{revtex4-2}

\usepackage{graphicx}
\usepackage{dcolumn}
\usepackage{bm}
\usepackage{hyperref}
\hypersetup{colorlinks=true,linkcolor=blue,citecolor=blue,filecolor=blue,urlcolor=blue}
\usepackage{color}
\usepackage{xcolor,soul}
\usepackage{alltt}
\usepackage{braket}
\usepackage{algcompatible}
\usepackage{newfloat}

\usepackage[obeyFinal,textsize=tiny, textwidth=\marginparwidth]{todonotes}
\presetkeys{todonotes}{fancyline, color=blue!30}{}

\newcommand{\figrefsub}[2]{\ref{#1}\hyperref[#1]{#2}}

\newcommand{\BKT}{{\textrm{BKT}}}

\renewcommand{\vec}[1]{\mathbf{#1}}
\newcommand{\svec}[1]{\boldsymbol{#1}}

\newcommand{\diff}{\textrm{d}}

\newcommand{\parderi}[2]{\frac{\partial{#1}}{\partial{#2}}}
\newcommand{\deri}[2]{\frac{\diff{#1}}{\diff{#2}}}
\newcommand{\abs}[1]{\left|#1\right|}

\newcommand{\mean}[1]{\left\langle #1 \right\rangle}

\newcommand{\kT}{T}
\newcommand{\al}{\alpha}
\newcommand{\om}{{\omega}}
\newcommand{\te}{{\theta}}
\newcommand{\ee}{\vec{e}}
\newcommand{\qq}{\vec{q}}
\newcommand{\rr}{\vec{r}}
\newcommand{\vs}{\vec{s}}
\newcommand{\vsi}{\vs_i}
\newcommand{\pp}{\vec{p}}
\newcommand{\mm}{\vec{m}}

\newcommand{\mmq}{\mm_{\qq}}
\newcommand{\mparq}{m_{\parallel,\qq}}
\newcommand{\mperpq}{m_{\perp,\qq}}
\newcommand{\wq}{w_{\qq}}
\newcommand{\nab}{\svec{\nabla}}

\newcommand{\chimq}{\chi_{m}(q)}
\newcommand{\chimparq}{\chi_{m\parallel}(q)}
\newcommand{\chimperpq}{\chi_{m\perp}(q)}

\newcommand{\nruns}{n_{\textrm{runs}}}
\newcommand{\eff}{_{\textrm{eff}}}
\newcommand{\Cminc}{C_m^{\textrm{inc}}}
\newcommand{\figit}[1]{$(#1)$} 

\newcommand{\inv}{^{-1}}
\newcommand{\transp}{^{\intercal}}

\usepackage{xargs}

\begin{document}

\author{Thomas Bissinger}
\email{thomas.bissinger@uni-konstanz.de}
\affiliation{Fachbereich Physik, Universit\"at Konstanz, 78457 Konstanz, Germany}

\author{Matthias Fuchs}
\email{matthias.fuchs@uni-konstanz.de}
\affiliation{Fachbereich Physik, Universit\"at Konstanz, 78457 Konstanz, Germany}

\date{\today}

\title{The BKT Transition and its Dynamics in a Spin Fluid}

\begin{abstract}
	We study the effect of particle mobility on phase transitions in a spin fluid in two dimensions. The presence of a phase transition of the BKT universality class is shown in an off-lattice model of particles  with purely repulsive interaction employing computer simulations. A critical spin wave region $0 < T < T_{\BKT}$ is found with a non-universal exponent $\eta(T)$ that follows the shape suggested by BKT theory, including a critical value consistent with $\eta_\BKT = 1/4$.  One can observe a transition from power-law decay to exponential decay in the static correlation functions at the transition temperature $T_\BKT$, which is supported by finite-size scaling analysis. A critical temperature $T_{\BKT} = 0.17(1)$ is suggested. Investigations into the dynamic aspects of the phase transition are carried out. The short-time behavior of the incoherent spin autocorrelation function agrees with the Nelson-Fisher prediction, whereas the long-time behavior differs from the finite-size scaling known for the static XY model. Analysis of coherent spin wave dynamics shows that the spin wave peak is a propagating mode that can be reasonably well fitted by hydrodynamic theory. The mobility of the particles strongly enhances damping of the spin waves, but the model lies still within the dynamic universality class of the standard XY model.
	
\end{abstract}

\maketitle

\section{Introduction}
\label{s:Intro}
The Berezinskii-Kosterlitz-Thouless (BKT) transition \cite{Berezinskii1971destruction,Kosterlitz1974,Kosterlitz2016kosterlitz} is ubiquitous in two-dimensional (2D) systems with a continuous symmetry. The symmetry cannot be broken in a conventional sense due to the Mermin-Wagner theorem, \cite{Mermin1966} i.e. the mean order parameter must vanish at all finite temperatures. Phase transitions driven by topological excitations (like vortices for magnetic systems\cite{Villain1974} or lattice defects for crystals\cite{Halperin1978}) circumvent this constraint. The excitations are bound at low temperatures, renormalizing the coupling constants of the system, and become unbound at a transition temperature $T_\BKT$.\cite{Kosterlitz1972} A characteristic feature of this type of transition is the spatial decay of order parameter fluctuations, which changes from a power-law behavior at all low temperatures to an exponential one above $T_\BKT$. In the language of critical phenomena, this is described by a continuous line of critical points in $0 < T \le T_\BKT$. When crossing $T_{\BKT}$, the stiffness constant $K$ of the effective Gaussian Hamiltonian displays a jump value from $K_\BKT = 2/\pi$ to zero. \cite{Kosterlitz1972,Nelson1977universal}

A paragon example of this behavior is the classical XY model of planar rotators on a 2D lattice, also known as the $O(2)$ model. In fact, the model goes by a lot of names and some authors distinguish the rotator model from the XY model as a quantum model for an easy-plane Heisenberg ferromagnet with the typical spin algebra.\cite{NelsonFisher1977,Evertz1996} We will use the term XY model synonymous with the rotator model,\cite{Chamati2006} the key feature is the symmetry of the relevant order parameter. Kosterlitz theoretically predicted the $\BKT$ scenario for XY systems.\cite{Kosterlitz1974,Kosterlitz2016kosterlitz} Further theoretical investigations and simulations confirmed this prediction.\cite{Tobochnik1979,Kogut1986,Ueda2021} The Mermin-Wagner theorem only holds in the thermodynamic limit of an infinitely extended system, as only there infinitely long-ranged Nambu-Goldstone modes suppress global order. Concurrently, the $\BKT$ transition in the XY model shows strong finite-size scaling, which was analyzed by an RG treatment and compared to simulation and experiment by Bramwell, Holdsworth \textit{et al.}\cite{Bramwell1993,Bramwell1994magnetization,Taroni2008} They predicted finite-size scaling of the magnetization with a universal finite-size exponent $\beta \approx 0.23$, which is in agreement with experimental observation.\cite{Wildes_1998}

These early results on the static properties of the classical XY model were soon supplemented by discussions of the dynamics. Villain \cite{Villain1974} put forward an approach to studying the hydrodynamics of spin waves on which Nelson and Fisher\cite{NelsonFisher1977} based their ``fixed-length`` hydrodynamic theory of dynamical critical scaling for easy-plane Heisenberg magnets. They predict divergent spin-wave peaks, a projection operator approach due to Menezes \textit{et al.}\cite{Menezes1993} arrives at a similar conclusion. The Nelson-Fisher predictions are also partially supported by works by Lepri and Ruffo\cite{Lepri2001}, who confirmed their short-time validity but also discovered further universal finite-size features in a rotator model.  Mertens \textit{et al.}\cite{Mertens1989} also investigated the spin dynamics of easy-plane Heisenberg magnets and predicted spin waves with a central peak and a propagating peak at low temperature that features a finite damping in the $q \to 0$ limit. In an extensive numerical study, Evertz and Landau \cite{Evertz1996} analyzed the neutron scattering function of the XY model, confirming the short-time prediction of the Nelson-Fisher approach while discovering structure beyond a single spin wave peak, including a central peak and fine-structure which they hypothesized to be due to scattering of multiple spin waves.

Mobile XY (MXY) models, also called spin fluids, provide off-lattice counterparts to the standard XY model discussed above.\cite{Wilding1996,Omelyan2001} They have been employed as phenomenological classical models to describe liquid-vapor interfaces \cite{Omelyan2009} or the effect of phonon-magnon couplings on the dynamics of bcc iron\cite{Perera2017}. With the current interest in critical phenomena of active matter\cite{Ramaswamy2010,Ginelli2015}, such systems are also interesting as inactive counterparts to models for living systems like the famed Vicsek model\cite{Vicsek1995}. Recently, spin-fluids were studied by Casiulis \textit{et al.} \cite{Casiulis2019,Casiulis2020Velocity,Casiulis2020Order} They found a ferromagnetism-induced phase separation (FIPS) replacing the BKT scenario. Beside the lattice-free spin fluids, lattice-gas generalizations of the XY model have also been studied, and BKT transitions were observed in lattice simulations.\cite{Chamati2006}

In this paper, we argue that in an equilibrium 2D MXY model with purely repulsive interaction, no phase separation occurs at the transition temperature and the system undergoes a $\BKT$-type transition at a critical temperature $T_{\BKT}$. To that end, we employ microcanonical MD simulations up to $N = (256)^2$ particles. We compare data of a mobile model to that of a similar model with disordered yet fixed particle positions, which we call a disordered XY (DXY) model. We will investigate the static properties of these models as well as their dynamical features, which are affected by the mobility of the spins.

In Section \ref{s:Models}, we present the mobile XY model, discussing some key properties, including the interaction potential, as well as introducing the thermodynamic quantities we use for the analysis. In Section \ref{s:Sim}, we give a short overview over the simulation parameters and some details of the implementation, then the results of the simulation are presented in Section \ref{s:Results}. We will consider the properties of the total magnetization and its susceptibility in Section \ref{s:Magnetization}, the effect of structural order in Section \ref{s:PositionalCorrelations}, then study the effect of finite size on static spin correlation functions in real and reciprocal space in Sections \ref{s:SpinCorrRealSpace} and \ref{s:SpinCorrReciprocal}, respectively. Considering the dynamic properties of the system, we compare data for the incoherent spin autocorrelation function to the Nelson-Fisher\cite{NelsonFisher1977} and Lepri-Ruffo\cite{Lepri2001} results in Section \ref{s:IncoherentSpinACF}. Finally, we study the coherent spin autocorrelation function in Section \ref{s:CoherentSpinACF} and their compatibility with a damped oscillator fit to the spin wave peak. Concluding remarks are given in Section \ref{s:Conclusion}.

\section{Definitions}
\label{s:Models}
This section summarizes the main definitions used throughout the paper. We introduce the model and then discuss various quantities that are relevant in the study of the critical phenomena. We conclude with a short discussion regarding the distinction between longitudinal and transversal fluctuations.

\subsection{The MXY and the DXY Model}
\label{s:Models:Model}
Our main focus is on the MXY model, or spin fluid\cite{Wilding1996,Casiulis2019}, with Hamiltonian
\begin{equation}
	\begin{aligned}
		H &= \sum_{j} \frac{\om_j^2}{2I} - \sum_{j\neq k} J(r_{jk})\cos(\te_{jk}) \\
		&\quad + \sum_{j} \frac{p_j^2}{2m} + \sum_{j\neq k} U(r_{jk}).
	\end{aligned}
	\label{eq:MXY}
\end{equation}
This describes $N$ isotropic particles with positions and momenta $\set{\rr_j,\pp_j}$ (bottom line) in 2D supplemented by a spin interaction depending on the interparticle distance. Spins are two-dimensional unit vectors $\vs_i = (\cos(\te_i),\sin(\te_i))\transp$ with angle $-\pi < \te_i \le \pi$. We use the shorthand $\rr_{jk} = \rr_k - \rr_j$ and similarly for $\te_{jk}$, and $r_{jk} = \abs{\rr_{jk}}$. After additionally defining $U_{jk} = U(r_{jk})$ and $J_{jk} = J(r_{jk})$, we can write the resulting equations of motion as
\begin{equation}
	\begin{aligned}
		&\deri{}{t}\te_j = \frac{\om_j}{I}, \quad
		\deri{}{t}\rr_j = \frac{\pp_j}{m},\\
		&\deri{}{t}\om_j = \sum_{k \neq j} J_{jk}\sin(\te_{jk})\\ 
		&\deri{}{t}\pp_j = -\sum_{k \neq j} \nab_{\rr_j}\left[U_{jk} - J_{jk}\cos(\te_{jk})\right].
	\end{aligned}
	\label{eq:MXY_EOM}
\end{equation}
We choose soft interactions for the two interaction potentials $J(r)$ and $U(r)$, giving them the form
\begin{equation}
	\begin{aligned}
		J(r) &= J_0(\sigma-r)^2\Theta(1 - r/\sigma),\\
		U(r) &= U_0(\sigma-r)^2\Theta(1 - r/\sigma),
	\end{aligned}
	\label{eq:JandU}
\end{equation}
with energy scales $J_0$ and $U_0$, an interaction range $\sigma$ and the Heaviside function $\Theta$. Throughout this paper, dimensionless quantities are used, which we obtain by choosing scales in the system such that we can set $m = 1$, $J_0 = 1$ and $\sigma = 1$ in \eqref{eq:MXY}, \eqref{eq:MXY_EOM} and \eqref{eq:JandU}. The remaining parameters of the system are then the particle number $N$, the (dimensionless) density $\rho=N \sigma^2/L^2$, the ratio $U_0 / J_0$ and the coefficient $I/(m\sigma^2)$. For our simulations, we will use various $N$ at fixed $\rho = 2.99$, and set $U_0/J_0 = 4$ as well as $I / (m\sigma^2) = 1$. Temperatures will be given in units of energy as well, that is $k_B = 1$.

Note an important difference between the interaction potentials \eqref{eq:JandU} and those of the spin fluids and Hamiltonian polar particles recently studied by Casiulis \textit{et al.}Casiulis \textit{et al.}\cite{Bore2016,Casiulis2019,Casiulis2020Velocity,Casiulis2020Order} While having the same $J(r)$, they chose $U(r) = U_0(\sigma-r)^4\Theta(\sigma - r)$. In our case, the overall interaction potential, $U(r) - J(r) \cos(\te)$, is repulsive everywhere for all choices of $r$ and $\te$ because of the shared square power $(\sigma - r)^2$ in the spatial dependence in \eqref{eq:JandU}. In contrast, Casiulis \textit{et al.} have an attractive region at $r$ close to $\sigma$ for spins with $\cos(\te) > 0$. This difference has important consequences for the model: While Casiulis \textit{et al.} find the system dominated by a melting transition that suppresses the BKT transition, we find BKT physics in the absence of melting.

We are particularly interested in the effect of mobility on the dynamic aspects of the phase transition. To isolate the effect of spatial disorder of the particles, we will sometimes work with what we call a disordered XY model (DXY), obtained by equilibrating the MXY model and then freezing the particles' positions and only allowing for the spin dynamics of the Hamiltonian \eqref{eq:MXY}.

\subsection{Basic Diagnostic Tools}
\label{s:Models:Tools}
To analyze the static properties, the standard order parameter is the magnetization per particle,
\begin{equation}
	\vec{m} = \frac{1}{N} \sum_{j} \vs_j
	\label{eq:Magnetization}
\end{equation}
and we will be interested in its modulus $m = \abs{\mm}$. Associated to $m$ is a susceptibility per spin\cite{Archambault1997magnetic}
\begin{equation}
	\chi_{m} = \beta N \mean{m^2 - \mean{m}^2} = \beta N \sigma_m^2
	\label{eq:chiabsm}
\end{equation}
with the variance $\sigma_m^2$. Equally interesting is the maximum value of the variance
\begin{equation}
	\sigma_{\max}^2 = \max_{T} \sigma_{m}^2
	\label{eq:sigmamax}
\end{equation}
Both $\chi_m$ and $\sigma_{\max}^2$ exhibit useful finite-size properties that are deemed characteristic of the BKT transition.\cite{Archambault1997magnetic}

\begin{figure*}[htb]
	\centering
	\includegraphics[width=.6\linewidth]{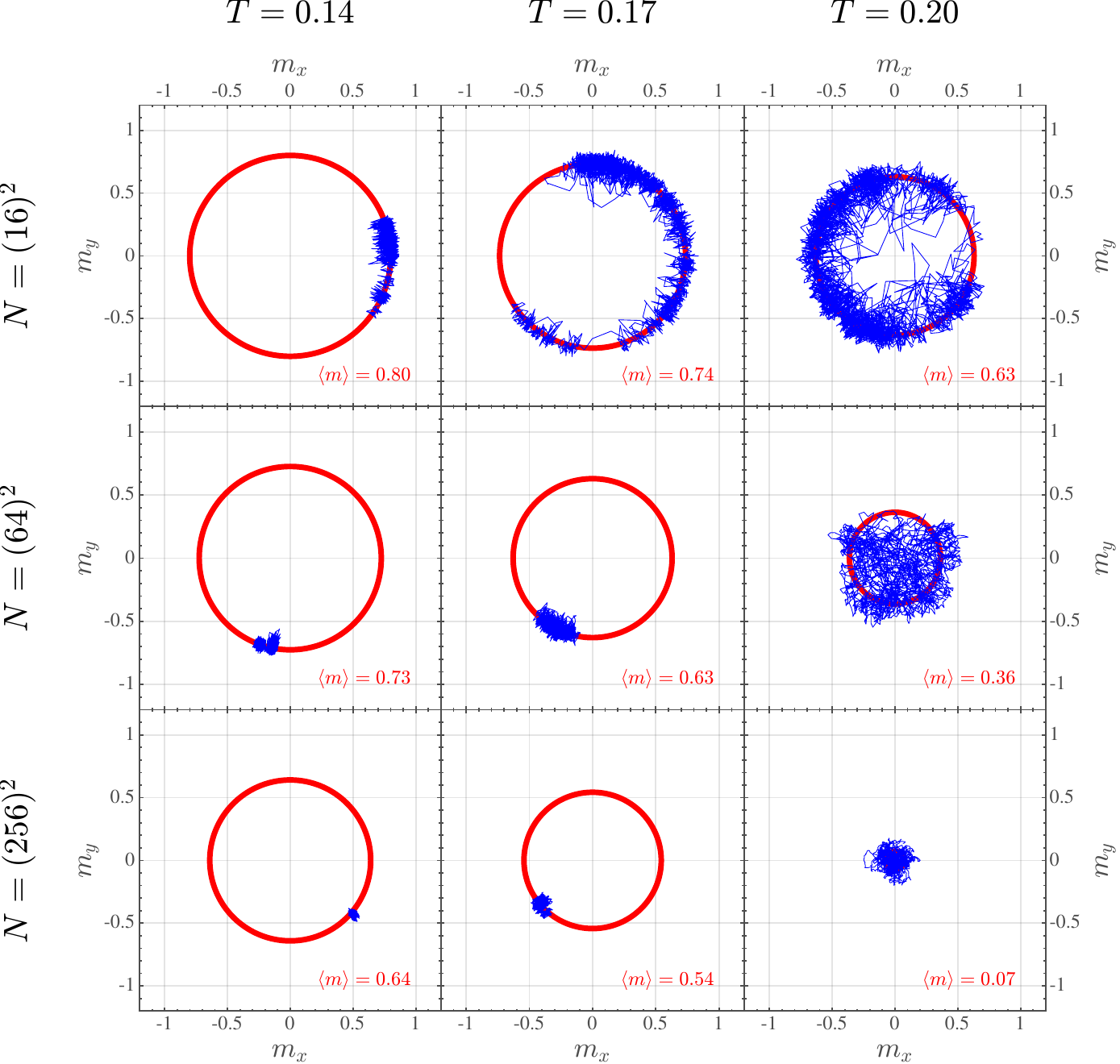}
	\caption{The finite-size magnetization below, at and above the transition in an MXY model. Evolution for a simulation time over $t_{\max} = 3\cdot 10^3$. Temperature increases from left to right, system size increases from top to bottom.}
	\label{fig:Magnetization_FS_Rings}
\end{figure*}

The spin stiffness $K$ is related to the helicity modulus $\Upsilon$ via $K = \beta \Upsilon$. $\Upsilon$ describes the elasticity of the spin alignment to twist distortions. Explicit formulae for $\Upsilon$ can be derived from the response of the system's free energy to a twist field,\cite{Sandvik2010computational} cf. Appendix~\ref{c:AppendixHelicity}. For the MXY model, the standard relation\cite{Hsieh2013}
\begin{equation}
	\Upsilon
	= \frac{1}{2A}\left(\mean{H_x + H_y} - \beta \mean{I_x^2 + I_y^2} \right).
	\label{eq:Upsilon}
\end{equation}
with $A = L_x L_y = L^2$ can be generalized according to the definitions\cite{BissingerThesis2022}
\begin{equation}
	\begin{aligned}
		H_x &= \frac{1}{2}\sum_{i \neq j } J(r_{ij})\cos(\te_{ij}) x_{ij}^2,\\
		I_x &= \frac{1}{2}\sum_{i \neq j } J(r_{ij})\sin(\te_{ij}) x_{ij},
	\end{aligned}
	\label{eq:H_x,I_X^2}
\end{equation}
with $x_{ij} = \rr_{ij} \cdot \ee_x$ the $x$-component of the distance vector $\rr_{ij}$. Analogous relations hold for $H_y$ and $I_y$. \cite{BissingerThesis2022}

We will further consider the magnetization and spin momentum densities
\begin{equation}
	\begin{aligned}
		\mm(\rr) &= \frac{1}{\sqrt{N}} \sum_j \vs_j \delta(\rr - \rr_j),\\
		w(\rr) &= \frac{1}{\sqrt{N}} \sum_j \omega_j \delta(\rr - \rr_j)
	\end{aligned}
\end{equation}
and their collective fluctuations
\begin{equation}
	\begin{aligned}
		\delta\mmq &= \frac{1}{\sqrt{N}}\sum_{j} (\vs_j - \mm) e^{-i\qq\cdot\rr_j},\\
		\delta\wq &= \frac{1}{\sqrt{N}}\sum_{j} \om_j e^{-i\qq\cdot\rr_j}.
	\end{aligned}
	\label{eq:mq,wq}
\end{equation}
Note that $\mm(\rr)$ has a different normalization than $\mm$.

\subsection{Longitudinal and Transversal Magnetization Fluctuations}
\label{s:Models:Fluctuations}
In the definition \eqref{eq:mq,wq}, we encounter a peculiarity of the finite size behavior of XY type systems. While in an infinitely expanded system, $\mean{m} = 0$ by the Mermin-Wagner theorem (which holds true for mobile particles as well\cite{BissingerThesis2022}), finite systems exhibit a well-defined value of $\mean{m} \neq 0$ at low temperatures. This is readily seen in Figure~\ref{fig:Magnetization_FS_Rings} for the MXY model, and similar results have been reported for the standard XY model.\cite{Archambault1997magnetic,CerrutiSola2000} At low temperatures, the modulus of the magnetization fluctuates around an average value, while its orientation diffuses over the circle with time. Increasing the system size leads to smaller fluctuations around the mean value and a reduced diffusion over the circle. At high temperatures, disorder sets in and the magnetization values are no longer bound to a circle. 

This means that, effectively, one has to treat finite XY systems as if they underwent symmetry breaking to an ordered state. We incorporate this by subtracting the spontaneous magnetization $\mm$ in the definition of $\delta\mmq$ in \eqref{eq:mq,wq}. Note that subtracting $\mean{\mm}$ is not sensible, as $\mean{\mm} = 0$ due to the degeneracy of the ``broken symmetry`` state. One could instead subtract $\mean{\mm}_0$ for some non-ergodic average $\mean{\cdots}_0$ that averages only over times where the orientation of $\mm$ does not change considerably, but we found that this does not yield different results than the ones obtained from \eqref{eq:mq,wq}. 

The infinite-system critical behavior can be extracted from the fluctuations transversal to the spontaneous order of the system. We will discuss this in more detail later, for now consider this a motivation to define
\begin{equation}
	\begin{aligned}
		\delta\mparq &= \delta\mmq \cdot \hat{\mm}, \qquad
		\delta\mperpq = \delta\mmq \cdot \hat{\mm}^\perp,
	\end{aligned}
\end{equation}
where the hat marks a unit vector $\mm = m \hat{\mm}$ and the orthogonal vector $\mm^\perp = (-m_y,m_x)$ is $\mm$ rotated by 90$^\circ$. Note that the $\vsi$ are not vectors with respect to spatial transformations, so there is no use in splitting into components parallel and perpendicular to $\qq$. This is a consequence of the spin orientation degree of freedom being independent of rotations of the particle position in \eqref{eq:MXY}.

\subsection{Correlation Functions}
With these definitions, let us consider the relevant correlation functions in our system. For spatial order, we are interested in the pair distribution function
\begin{equation}
	\rho g(r) = \frac{1}{N} \sum_{i \neq j}\mean{\delta(r - r_{ij})},
	\label{eq:g(r)}
\end{equation}
$\rho$ being the particle density. We will also consider the spin-spin correlation function
\begin{equation}
	\rho g(r) C_m(r) = \frac{1}{N}\mean{\sum_{i \neq j} \vs_i\cdot \vs_j \delta(r - r_{ij})},
	\label{eq:C_m(r)}
\end{equation}
which measures the correlations of a spin's orientation to that of another spin a distance $\rr$ apart. The factor $\rho g(r)$ is scaled out, it accounts for the packing of particles. It is known from Kosterlitz's BKT theory of the standard XY model\cite{Kosterlitz1974} that the behavior of $C_m(r)$ changes from exponential decay to power-law decay at the transition temperature, and we will see how that transfers to the MXY model in Section \ref{s:SpinCorrRealSpace}.

We can define $q$-dependent susceptibilities
\begin{equation}
	\begin{aligned}
		\chimq &= \mean{\abs{\delta\mmq}^2} = \chimparq + \chimperpq, \\
		\chimperpq &= \mean{\abs{\delta\mperpq}^2},\\
		\chimparq &= \mean{\abs{\delta\mparq}^2}.
	\end{aligned}
	\label{eq:chimq}
\end{equation}
The susceptibility $\chimq$ is related to the Fourier spectrum of $C_m(r)$ via \cite{BissingerThesis2022}
\begin{equation}
	\begin{aligned}
		&\rho g(r) \left(C_m(r) - \mean{m^2}\right)\\
		&= \int \frac{\diff^d q}{(2\pi)^d} e^{i\qq\cdot \rr} \left[ \chi_m(q) - \left(1 - \mean{m^2}\right) \right].
	\end{aligned}
\end{equation}
The susceptibility $\chi_w(q)$ follows from equipartition, $\mean{\om_i\om_j}=\kT\delta_{ij}$, it is $\chi_w(q) = \mean{\abs{\wq}} = \kT$.

A characteristic quantity of the dynamic $\BKT$ universality class is the incoherent spin autocorrelation function
\begin{equation}
	\Cminc(t) = \mean{\vs_i(0)\cdot\vs_i(t)},
	\label{eq:C^inc(t)}
\end{equation}
which compares the orientation of a spin to its orientation after a time $t$ has passed. Note that in the MXY model, the spin $i$ will change position with $t$.

Finally, to study the collective dynamics of the system, we shall study the spin-spin time correlation function
\begin{equation}
	C_{m}(q,t) = \mean{\delta\mmq \cdot \delta\mmq(t)}
	\label{eq:C_m(q,t)}
\end{equation} 
and the corresponding $C_{m\perp}(q,t)$, $C_{m\parallel}(q,t)$ and $C_{w}(q,t)$ as well as the resulting power spectra
\begin{equation}
	S_{m}(q,\omega) = \int_{-\infty}^\infty \diff t\ C_{m}(q,t) e^{i\om t}
	\label{eq:S_m(q,om)}
\end{equation}
alongside $S_{m\perp}(q,\omega)$, $S_{m\parallel}(q,\omega)$ and $S_{w}(q,\omega)$.

\section{Simulation}
\label{s:Sim}
We employ MD simulations of our system. Our simulations of the MXY model describe $N \in \set{(16)^2,(32)^2,\ldots,(256)^2}$ particles in a square area of linear dimension $L \in \set{9.25,18.5,37,74,148}$ with periodic boundary conditions. The box lengths yield a density of $\rho = N/L^2 = 2.99$. This value was chosen in part to be comparable to results by Casiulis \textit{et al.}\cite{Casiulis2019} and to ensure a low temperature crystallization of the MXY model. We simulate the models at discrete temperatures in a range of $0.01 \le T \le 0.51$, with special interest in data points around the transition temperature $T \approx 0.17$ (see below). As discussed in Section~\ref{s:Models:Model}, we set the scales of the simulation by $m=J_0=\sigma=1$ and fix the remaining free parameters to $U_0/J_0=4$ and $I/m\sigma^2 = 1$.

The dynamics is simulated by a velocity Verlet algorithm with time step $\delta t = 10^{-2}$ in the reduced time units. For equilibration, we start from a disordered system and use annealing to cool the system, where after each $10$ time units all momenta are multiplied by a factor of $\xi = 0.999$. An effective way of reaching bigger system sizes appeared to be to use fourfold copies of a smaller system, say $N=(128)^2$ particles, as initial conditions for the equilibration of the larger system of $N=(256)^2$ particles, joining the copies at the periodic boundary. The approach was effective for the mobile model. For a static model with comparable parameter choice, equilibration was much slower and this approach does not work.\cite{BissingerThesis2022}

Our criterion for equilibration was to calculate the correlation functions $C_{m\perp}(q,t - t_0)$ starting at different times $t_0$ and averaging them over $n_{\textrm{runs}} = 125$ runs. We consider a system to be equilibrated when $C_{m\perp}(q,t - t_0)$ is invariant with respect to different starting times $t_0$ for all finite $\qq$ compatible with the system dimensions.

After equilibration, annealing is switched off and we perform microcanonical simulations over up to $10^4$ time units to sample static and dynamic quantities. A preliminary simulation study\cite{Hoefler2019Berez-4733} tested that our choice of time step is sufficiently accurate to have only a negligible energy drift over the simulation time. For each pair of temperature and system size, averages are performed over $\nruns = 500$ runs. Per run, static quantities are computed $100$ times (every $100$ time units) and averaged, while time-correlation functions are computed once. Depending on the correlation time, the statistical error on the time-correlation functions is larger than that of the static quantities. Space-dependent quantities are obtained by a binning scheme, for $q$-dependent quantities, all wave vectors $\qq$ are considered individually. Note that due to the microcanonical nature of the simulations, we have no direct control over the temperature $T$. However, since most data for similar investigations is given with respect to a temperature $T$, we will do so, too, instead of giving results with respect to the conserved energy $E$.

For the DXY model, we start out from an equilibrated geometry of the MXY model and freeze all particle motion. We let the system equilibrate with the same annealing scheme and for at least $4\cdot10^3$ time units, after that averages are computed in the same way.

To gain time correlation functions, we calculate $C_{m\perp}(q_i,t_j)$ for a selection of admissible wave vectors $q_i$ and times $t_j$ and average them over the $\nruns = 500$ runs. The quantities $S_{m\perp}(q_i,\om_j)$ are then calculated using the FFT algorithm. For long times, the $C(q,t)$ obtained by this procedure will not decay but be dominated by fluctuations of the order $\nruns^{-1/2}$. To suppress their influence on the power spectra, we use Gaussian resolution functions $f(t) = \exp[-t^2/2\tau^2]$ with some time scale $\tau$ (the choice of which will be discussed in Section \ref{s:CoherentSpinACF}). For the Fourier transform, we then use 
\begin{equation}
	\tilde{S}_{m\perp}(q,\om) = {\cal F}[C_{m\perp}(q,\cdot)f(\cdot)](\omega).
	\label{eq:FourierTrafoWithConvolution}
\end{equation}
We checked that the resulting spectra do not depend strongly on the specific choice of $f(t)$ and that there are no qualitative improvements by increasing $\nruns$.

\section{Results}
\label{s:Results}

\subsection{Statics}
\label{s:SimStatic}
In this section, we study the static properties of the MXY model. We analyze the finite-size behavior of the mean magnetization $\mean{m}$ \eqref{eq:Magnetization} and the susceptibility $\chi_m$ \eqref{eq:chiabsm} as well as the correlation functions $C_m(r)$ \eqref{eq:C_m(r)} and $\chimq$ \eqref{eq:chimq} and their scaling behavior to obtain estimates of the transition temperatures and critical exponents.

It is worth noting that results for static properties of the MXY and the DXY model can hardly be distinguished. Differences are not systematic and lie within the error bars. This is not surprising: We obtain the DXY positions by freezing out the MXY spatial degrees of freedom. Thus, the positional order in the DXY model is the same as that of the MXY model, which supports the interpretation that the couplings $J_{jk} = J(\abs{\rr_k-\rr_j})$ in the DXY model are akin to (ferromagnetic) spin glass couplings with probability distribution given by the distribution of MXY particle separations $\abs{\rr_k-\rr_j}$.

For this reason, all the results presented in this section apply equally well for the MXY as the DXY model, and we show only data for the MXY model. We will discuss differences when addressing dynamics in Section \ref{s:SimDyn}.

\subsubsection{Absolute Magnetization and Finite Size Scaling}
\label{s:Magnetization}
\begin{figure}[tb]
	\centering
	\includegraphics{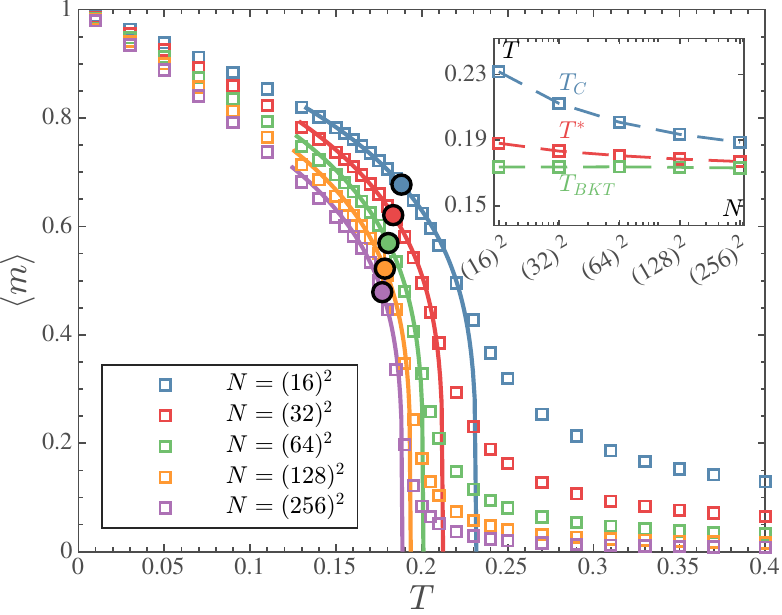}
	\caption{Magnetization for the MXY model. Squares are simulation data, solid lines are fits to \eqref{eq:BramwellHoldsworthFit}. Circles mark the temperatures $T^*$ defined by \eqref{eq:Tstar}. Inset: Scaling of finite-size critical temperatures, see text and Bramwell and Holdsworth\cite{Bramwell1994magnetization}. $\rho = 2.99$ (here and throughout).}
	\label{fig:MvsT}
\end{figure}
Figure~\ref{fig:MvsT} offers a first indication towards the presence of a BKT phase transition. It depicts the temperature dependence of the mean modulus of magnetization $\mean{m}$, \eqref{eq:Magnetization}, in the MXY model.

We observe a crossover between a low temperature regime with a linear decrease of magnetization with increasing temperature and a high temperature regime where the magnetization goes to zero with the system size. For $T \to \infty$, it reaches a plateau value at $\mean{m} = N^{-1/2} \sqrt{\pi}/2$, which is the mean magnetization for $N$ randomly oriented $2$-dimensional classical spins, a consequence of the multidimensional central limit theorem.\cite{BissingerThesis2022}

We will call the low temperature regime the spin wave phase. The crucial ingredient to a spin wave theory is the approximation of the spin-spin interaction by an effective harmonic Hamiltonian. Neither the amorphous arrangement of spin sites nor the resulting inhomogeneity of the coupling terms interfere with assuming $\cos(\te) \approx 1 - \frac{1}{2}\te^2$.  We can test the extent to which the Hamiltonian \eqref{eq:MXY} lends itself to a spin wave approximation by comparing to results from spin wave theory based on lattice XY models. One interesting property is the finite-size scaling behavior of the magnetization. The Hamiltonian becomes Gaussian in spin wave theory, which allows for the prediction\cite{Archambault1997magnetic,Tobochnik1979,Bramwell1993}
\begin{equation}
	\mean{m}(T,L) = (2N(L))^{-\eta(T)/4}.
	\label{eq:M_SW}
\end{equation}
The exponent $\eta$ is inversely proportional to the spin wave stiffness $K = 1/(2\pi\eta)$.\cite{Kosterlitz1974} Note that \eqref{eq:M_SW} was derived from a lattice-based spin wave theory of the XY model. Our model is off-lattice and replaces the relationship $N=L^2$ for a square lattice by $N = \rho L^2$. It is therefore entirely justified to replace $2N(L)$ by some factor $\alpha^2 \rho L^2$. We compared data for various $\alpha$ and found that the results below do not depend strongly on $\alpha$ (for large systems, $N=(256)^2$, the transition temperature $T_{\BKT}$ defined below varies by less than $5\%$ for $1 \le \alpha^2 \le 4$). For that reason, we keep to the literature value of $\alpha^2 = 2$ for the remainder of this study.

Equation \eqref{eq:M_SW} can be used to fit the $N$-dependent data over the whole temperature range $0.01 \le T \le 0.51$. Reasonable agreement is reached for temperatures $T \le 0.18$, with small deviations starting to appear around $T \approx 0.16$ (see below, Figure~\ref{fig:SpinCorrExponents}). Above $T = 0.19$, the magnetization does not scale as a clear power law in $N$, while at high temperatures, the random spin power law $\mean{m} \propto N^{-1/2}$ sets in (we do not show this data).

Equation \eqref{eq:M_SW} can not only be used to check the applicability of a spin wave approximation. It can also serve as a method to extract the exponent $\eta$ from the simulation data. We will return to this prescription when discussing Figure~\ref{fig:SpinCorrExponents} below.

In Kosterlitz's seminal paper, the presence of vortices renormalizes the naive stiffness constant $K_0$ to a new value $K$.\cite{Kosterlitz1974} When discussing finite systems, vortices can only be integrated out up to the extension of the system, leading to a size-dependence $K(L)$, and consequently size-dependence in $\eta(L)$. At temperatures well below the transition temperature $T_{\BKT}$ and for fixed system sizes $N$, the vortex separation $\xi_v < L$ (with $L = \sqrt{N/\rho}$), therefore $K(L) = K(\infty) = K$ and there is truly no size-dependence of the exponent in \eqref{eq:M_SW}. However, close to $T_{\BKT}$ and above, $\xi_v > L$ and finite-size effects round off the jump in the stiffness $K(L)$ and the suppression of $\mean{m}$ above $T_{\BKT}$. Bramwell and Holdsworth studied such phenomena with a finite-size renormalization group analysis.\cite{Bramwell1993,Bramwell1994magnetization} They found that the transition temperature $T_{\BKT}$ is to be replaced by a size-dependent $T^*(L)$ defined to be the temperature where the exponent $\eta$ reaches the critical value
\begin{equation}
	\eta(T^*(L),L) = 1/4.
	\label{eq:Tstar}
\end{equation}
For $T \gtrsim T^*$, $\mean{m}$ follows a power-law behavior
\begin{equation}
	\mean{m}(T,L) \sim (T_C(L) - T)^{\beta}
	\label{eq:BramwellHoldsworthFit}
\end{equation}
The additional critical temperature $T_C(L) > T^*(L)$ is the temperature where the correlation length $\xi$ of the spin-spin correlation function $C_m(r)$ \eqref{eq:C_m(r)} exceeds the system size coming from the high temperature disordered phase. The exponent $\beta$ is experimentally measurable with $\beta = 3\pi^2/128 \approx 0.23$.

The circles in Figure~\ref{fig:MvsT} mark temperatures $T^*$ and the corresponding magnetizations, they are obtained by linearly interpolating $\mean{m}(T,L)$ to find the value $T^*$ defined by \eqref{eq:Tstar}. The solid lines in Figure~\ref{fig:MvsT} are fitted using \eqref{eq:BramwellHoldsworthFit}. For the fit, we estimated $T_C$ by a least squares method in the range $0.13 \le T \le T_C$ with special emphasis on the region around $T^*$.

In the limit $N \to \infty$, both $T^*$ and $T_C$ should converge to the same value $T_{\BKT}$. In their RG analysis, Bramwell and Holdsworth\cite{Bramwell1993} derive the relation $T_{\BKT}(L) = (4 T_C(L) - T^*(L))/3$. The inset in Figure~\ref{fig:MvsT} shows the convergence of the three temperatures $T_C$, $T^*$ and $T_{\BKT}$. All curves seem to go smoothly and monotonously with $N$, with $T_C$ and $T^*$ decreasing, while $T_{\BKT}$ changes only slightly. At the largest system size, we find $T_{\BKT}(L_{\max}) = 0.173$, $T_C(L_{\max}) = 0.189$ and $T^*(L_{\max}) = 0.177$. Given that our simulations are microcanonical with a limited temperature control and that the analysis relies on fits to simulation data, we estimate the transition to lie in the range $T_{\BKT} = 0.17(1)$.

\begin{figure}[tb]
	\includegraphics{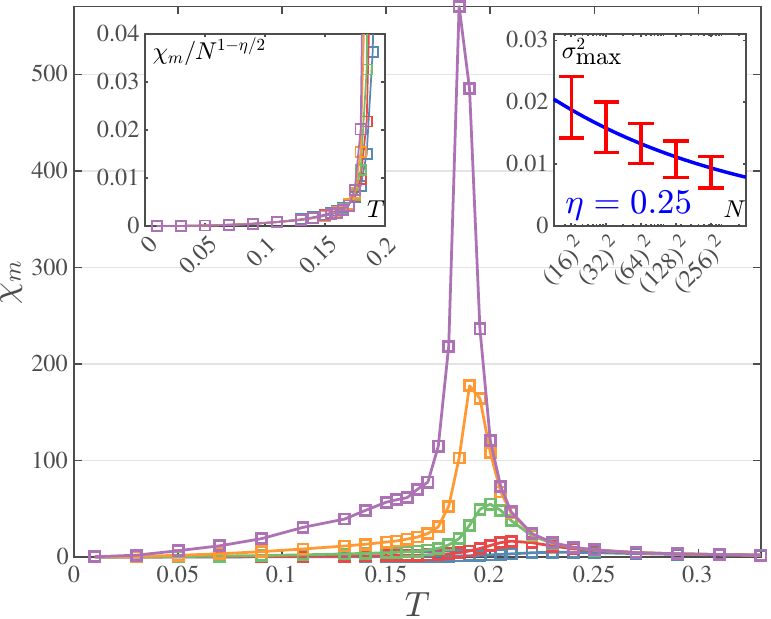}
	\caption{Behavior of the susceptibility $\chi_{m}$, defined in \eqref{eq:chiabsm}, for different $N$. Color code follows Figure~\ref{fig:MvsT}. Left inset: Spin wave scaling $\chi_m(T,L) \sim N^{1-\eta/2}$ below the transition. Right inset: Size-dependence of the maximum $\sigma_{\max}^2$, \eqref{eq:sigmamax}, with a power-law fit $\sigma_{\max}^2 \sim L^{-\eta}$ with $\eta = 1/4$. Simulation data red error bars, fit solid line.}
	\label{fig:AbsSusceptibility}
\end{figure}
Another indication of a finite-size scaling consistent with the BKT transition is the study of the susceptibility $\chi_{m}$ from \eqref{eq:chiabsm} and the maximum values of the variance $\sigma_{\max}^2$, \eqref{eq:sigmamax}. In Figure~\ref{fig:AbsSusceptibility}, the susceptibility shows more and more pronounced peaks with increasing system size, and indeed the peak location is consistent with $T_C$ from the analysis of the magnetization. It has been shown\cite{Archambault1997magnetic,Leoncini1998} that $\chi_m$ should scale as $\chi_{m}(L,T) \sim N^{1-\eta(T)/2}$, and consequently $\sigma_{\max}^2 \sim L^{-\eta}$. We obtained $\eta(T)$ from collapsing \eqref{eq:M_SW}, and the left inset of Figure~\ref{fig:AbsSusceptibility} demonstrates the accuracy of this scaling. For temperatures $T \ll T_{\BKT}$, the collapse works particularly well. When approaching the transition temperature, the fluctuations increase dramatically and simulations at different system sizes are not equally close to their respective transition temperatures $T_C$ and $T^*$. Thus, verifying the validity of the $\chi_m$ scaling becomes more difficult in the vicinity of $T_{\BKT}$. The right inset of Figure~\ref{fig:AbsSusceptibility} investigates the power-law behavior of $\sigma_{\max}^2 \sim L^{-\eta}$.\cite{Archambault1997magnetic} The data is well fitted for the choice $\eta = 1/4$. A best fit is obtained for $\eta_{\textrm{best}} = 0.28$, yet within the rather large error bars on $\sigma_{\max}^2$ visible in Figure~\ref{fig:AbsSusceptibility}, the exact value of $\eta$ cannot be determined with great accuracy.

\begin{figure}[tb]
	\centering
	\includegraphics{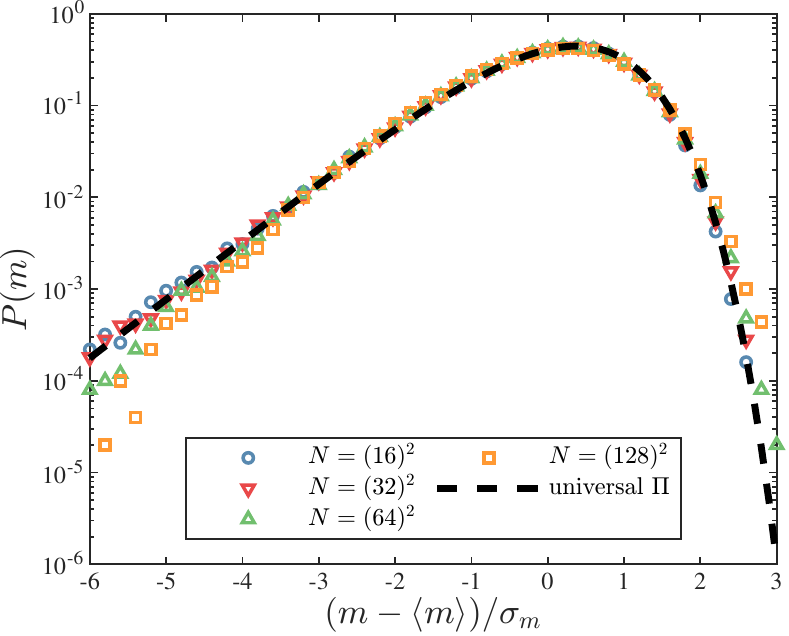}
	\caption{Finite-size analysis of the magnetization histogram for the MXY model close to critical temperature. The universal scaling curve $\Pi$ is taken from Archambault \textit{et al.}\cite{Archambault1997magnetic}. $\sigma_m$ is the variance of the magnetization, see \eqref{eq:chiabsm}. Data taken at $T = 0.17$.}
	\label{fig:Mhistogram}
\end{figure}
A universal form of the probability distribution function $P(m)$ of the modulus of the magnetization has been reported for the standard XY model and other highly correlated 2D systems.\cite{Archambault1997magnetic,Bramwell2000} Such a behavior should also be expected from the MXY model, and indeed we find universal behavior in Figure~\ref{fig:Mhistogram} at $T= 0.17$. Here, the variance $\sigma$ is related to the susceptibility by $\sigma^2 = (T/N)\chi_m$. Differences from the scaling functions in the tails of the distribution are most likely due to statistical errors in the data, since larger deviations become very rare events. 

As a closing remark to the discussion of the magnetization, the finiteness of the interval $T^* < T < T_C$  leads us to expect to find an anomalous transition region between the two temperatures that is no longer fully described by spin wave theory at all system sizes while nonetheless having system-wide magnetization correlations due to $\xi > L$.

\subsubsection{Positional Correlations}
\label{s:PositionalCorrelations}
\begin{figure}
	\includegraphics[width=.99\columnwidth]{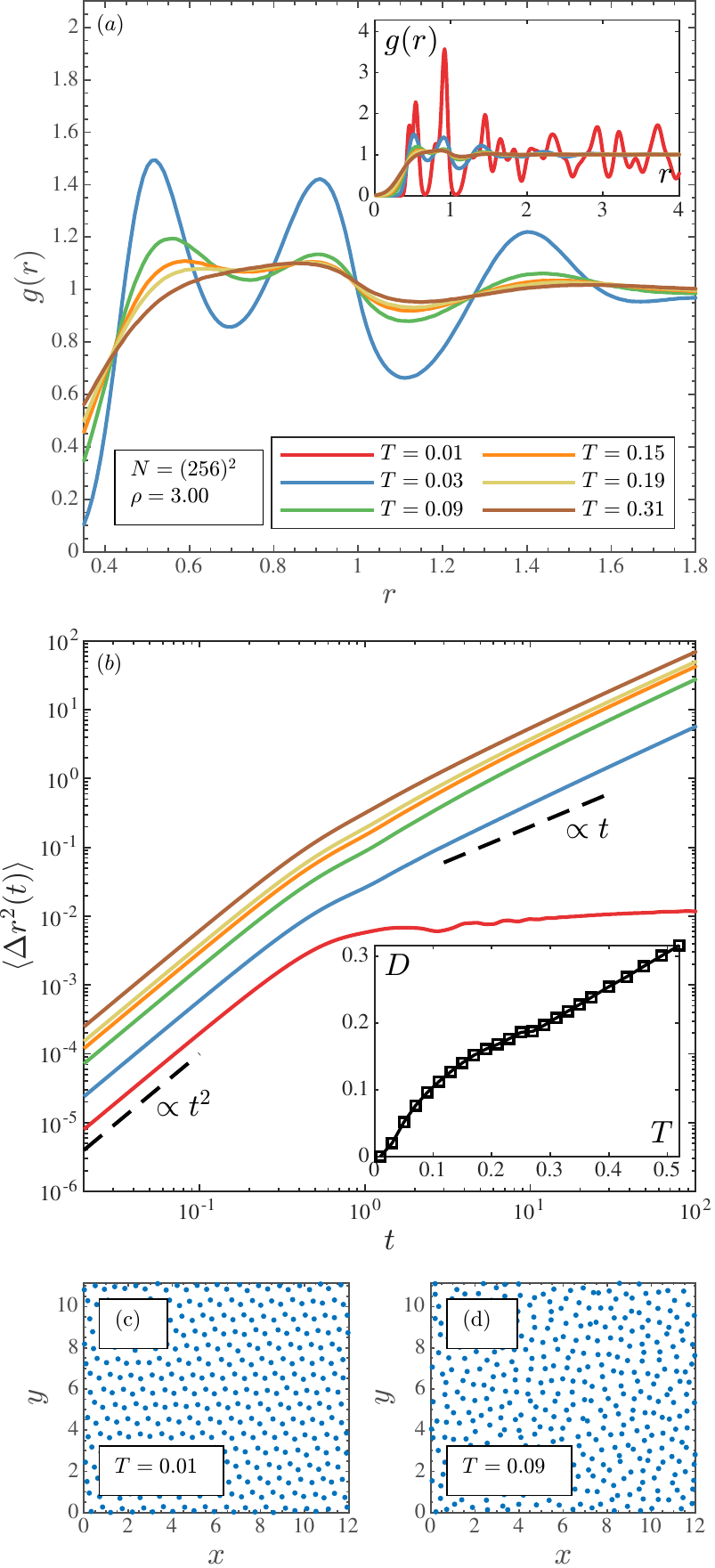}
	\caption{Structural properties of the MXY model. $N = (256)^{2}$, $\rho = 2.99$ in all images. \figit{a} Radial distribution function $g(r)$ at different temperatures. The inset includes the lowest temperature at $T= 0.01$. \figit{b} MSD of particle position $\mean{\Delta r^2(t)}$, the inset shows the long-time diffusion coefficient $\mean{\Delta r^2(t)} = 4Dt$. Legend see \figit{a}. \figit{c} and \figit{d} Particle position snapshot at $T = 0.01$ and at $T = 0.09$, respectively.}
	\label{fig:gr}
\end{figure}
In the standard XY model, positions are fixed to a lattice, but the MXY model allows for particle motion. Figure~\figrefsub{fig:gr}{\figit{a}} shows the radial distribution function $g(r)$ for different temperatures for systems with $N=(256)^2$ particles. As the inset shows, traces of crystalline order are only apparent at the lowest of temperatures, where the particles self-assemble into a lattice due to the density. Already at temperatures of $T = 0.03$, well below the transition temperature around $T_{\BKT} \approx 0.17$, the system is in a fluid phase, and there are no qualitative changes when passing through the magnetic transition. This leads us to the conclusion that the transition is due to the spin dynamics and not influenced by any liquid-solid phase transition.

The liquid behavior of the system becomes apparent when studying the mean-squared displacement (MSD) of the particles,
\begin{equation}
	\mean{\Delta r^2(t)} = \frac{1}{N}\sum_{j} \mean{(\rr_j(t) - \rr_j(0))^2}
	\label{eq:MSD}
\end{equation}
in Figure~\figrefsub{fig:gr}{\figit{b}}. At all temperatures, the MSD initially grows as the ballistic law $\mean{\Delta r^2(t)} = (2\kT/m) t^2$. For all but the lowest temperature, there is a transition to a long-time diffusive regime $\mean{\Delta r^2(t)} \sim t$. Freezing effects can only be observed at $T = 0.01$.

Figure~\figrefsub{fig:gr}{\figit{c}\figit{d}} present snapshots of a small subdomain of the system at temperatures $T=0.01$ and $T=0.09$, respectively. They further support our observation that the system crosses into a liquid-like phase well below the finite-size critical temperature domain. At the same time, we see that the zero-temperature structure is a stretched honeycomb lattice. This is in qualitative agreement with studies on the Hertzian model\cite{Pamies2009,Miller2011}, a spin-free soft sphere/disc model with the interaction potential $U(r)$ from \eqref{eq:JandU}, however with a dependence on $(1-r/\sigma)^{\alpha}$ with variable exponent $\al \neq 2$. In particular, Miller and Cacciuto \cite{Miller2011} found a stretched honeycomb lattice for $\al = 3/2$ at densities of $\rho=2.99$ and a reduced temperature $T_r = T_c/\epsilon \le 0.0125$ with the energy scale $\epsilon$. For the three values $\alpha \in \set{3/2,5/2,7/2}$ they studied, this was the highest transition temperature at the density $\rho = 2.99$. Above this temperature, all systems were in a liquid state. To compare to our data, note that at very low temperatures, when spins are almost aligned, we can assume a (dimensionless) energy scale $\epsilon/J_0 = U_0/J_0 - 1 = 3$ for the total interaction due to $U(r) - J(r)\cos(\te)$, cf. \eqref{eq:JandU}. A system well above the rough estimate $T_c = \epsilon T_r = 0.0375$ for the transition temperature should therefore exhibit liquid-like spatial ordering. This is a purely heuristic argument -- because of the different values of $\al$ and a lack of spin interaction in the Hertzian model, quantitative estimates would require further investigation.

\subsubsection{Spin Correlations in Real Space}
\label{s:SpinCorrRealSpace}
\begin{figure}[tb]
	\includegraphics{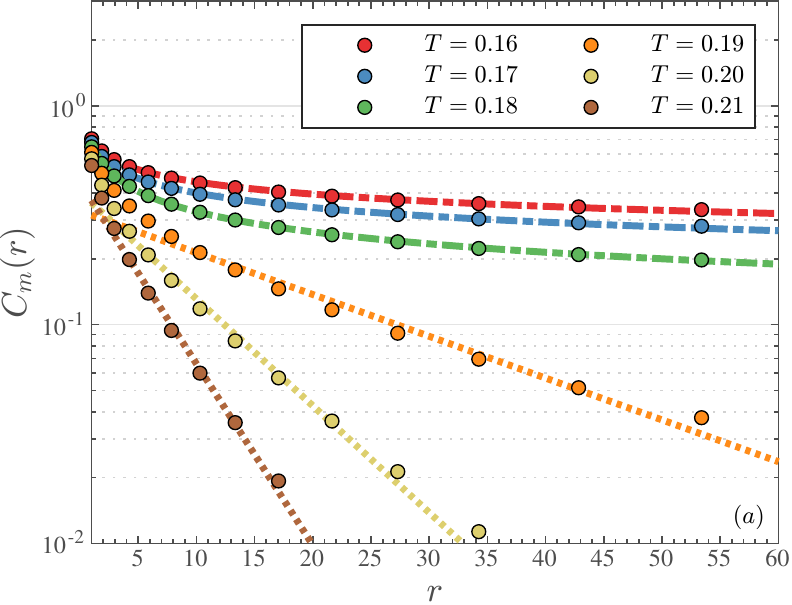}\\[.5em]
	\includegraphics{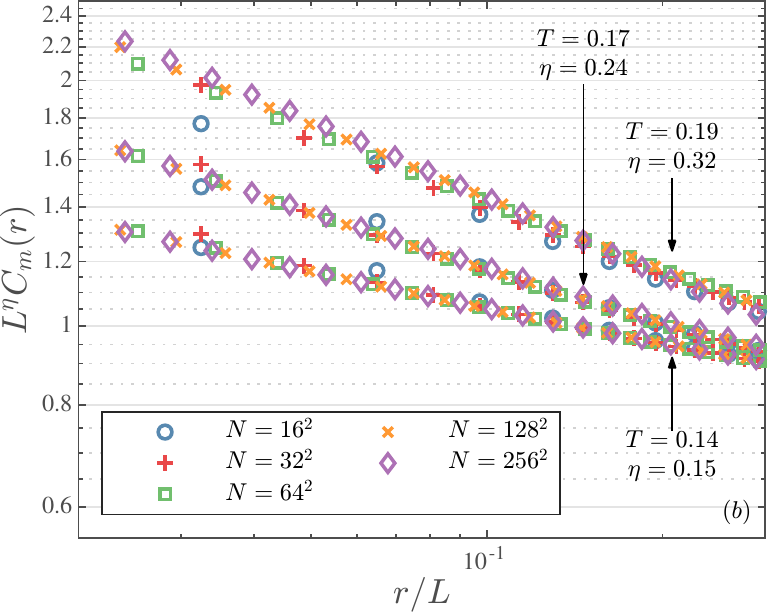}
	\caption{Spin correlation functions for MXY model. System size $N = (256)^2$ and density $\rho = 2.99$. The critical temperatures at this system size are $T_C = 0.19$ and $T^{*} = 0.17$. \figit{a} Raw data with fits. Points are simulation data, dashed lines are fits to a power-law decay with exponent $\eta(T)$, dash-dotted lines are fits to exponential decay with correlation length $\xi(T)$, cf. \eqref{eq:SpinCorrAsymptotic}. \figit{b} Scaled values for finite-size analysis according to \eqref{eq:SpinCorrScaling}.}
	\label{fig:SpinCorr}
\end{figure}
BKT-like finite-size scaling encourages us to venture further into the study of the BKT universality class of the mobile model. A most characteristic feature of BKT universality is the transition from paramagnetic short-range disorder at high temperatures to ferromagnetic quasi long-range order (QLRO) at temperatures below the transition.\cite{Kosterlitz1974} This transition is visible in the asymptotic behavior of the spin-spin correlations $C_m(r)$, \eqref{eq:C_m(r)},
\begin{equation}
	C_m(r)
	\sim \begin{cases}
		r^{-\eta(T, L)},\qquad & \textrm{for } T < T^*(L),\\
		e^{-r/\xi(T, L)},\qquad & \textrm{for } T > T_C(L).
	\end{cases}
	\label{eq:SpinCorrAsymptotic}
\end{equation}
For each pair of $T$ and $L$, we apply fits to \eqref{eq:SpinCorrAsymptotic} in the range $0.3 \le r \le 0.4 L$ (finite-size effects set in at $r \approx L/2$), which determine $\eta(T, L)$ and $\xi(T,L)$. Figure~\figrefsub{fig:SpinCorr}{\figit{a}} demonstrates the accuracy of these fits for various temperatures around the finite-size critical region at $N = (256)^2$. Standard errors in the simulation data are smaller than the symbols. At temperatures $T \ge 0.20$, exponential decay (dotted line) fits the data best. At temperatures $T \le 0.18$, the power-law fit (dashed line) gives the best agreement. There is an intermediate region around $T = 0.19$ (orange data points) where both fits are inaccurate. Finite-size effects are strongest there.

While this is a valid procedure for obtaining $\eta$ and $\xi$, it has some drawbacks. For once, there is some degree of size-dependence left in $C_m(r)$ and thus in $\eta$ and $\xi$. The size-dependence is strongest close to the transition. As mentioned above, in the region around the transition, neither functional form of $C_m(r)$ offers good agreement with the data. Finally, the asymptotic behavior $C_m(r) \to m^2$ with $r \to \infty$ does not decay to zero in finite systems, which is evident in the spin wave phase due to \eqref{eq:M_SW}.

To obtain stronger evidence of QLRO at low temperatures, we analyze $\eta$ in a region $T \lesssim T_{\BKT}$. In this region, finite-size scaling predicts\cite{Evertz1996}
\begin{equation}
	C_m(r, L)
	= L^{-\eta} f\left(\frac{r}{L}\right).
	\label{eq:SpinCorrScaling}
\end{equation}
Searching for an optimal fit to \eqref{eq:SpinCorrScaling} can thus provide a size-independent value for $\eta$. We used an error estimate for this kind of data collapse similar to the one proposed by Bhattacharjee and Seno\cite{Bhattacharjee2001} to obtain values for $\eta$ by minimization. Some examples of the scaling relation \eqref{eq:SpinCorrScaling} are plotted in Figure~\figrefsub{fig:SpinCorr}{\figit{b}}. We can see that data collapse works well for temperatures up to $T \approx 0.17$, which agrees with our estimate $T_{\BKT} \approx 0.173$, while deviations become more pronounced at higher temperatures, e.g. the optimal value of $\eta = 0.32$ at $T = 0.19$ yields a considerably poorer quality of data collapse. This is because for the large $N = (256)^2$ system, $C_m(r)$ is at its critical $T_C$ and follows an exponential decay, while smaller systems are below their respective $T_C$ and power-law behavior is still accurate there.

\subsubsection{Spin Correlations in Reciprocal Space}
\label{s:SpinCorrReciprocal}

The scaling result \eqref{eq:SpinCorrScaling} transfers to the Fourier transform of $C_m(r,L)$ as
\begin{equation}
	\chi_m(q, L)
	= L^{2-\eta} g(qL).
	\label{eq:ChiScaling}
\end{equation}

\begin{figure*}[htb]
	\includegraphics{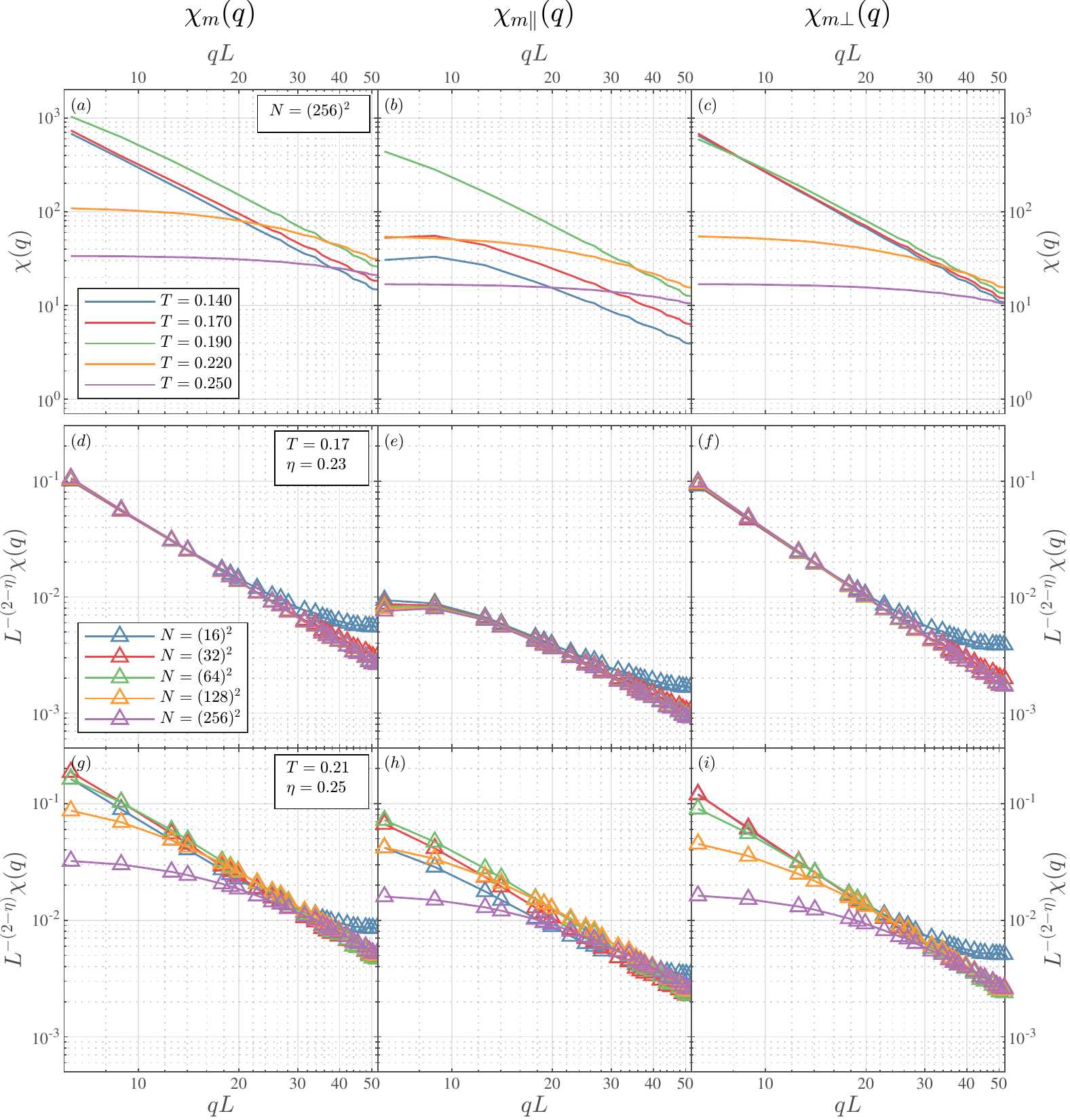}
	\caption{$q$ dependence of the total magnetization susceptibilities. From left to right, we compare the two contributions $\chi_{m\parallel}(q)$ and $\chi_{m\perp}(q)$ to the full $\chi_{m}(q)$. The top row shows the behavior at a fixed system size and different temperature, the bottom two rows exemplify the $q$-dependence for temperatures at and above the transition. Panels \figit{a} - \figit{c} Different temperatures at the fixed system size $N = (256)^2$, $L = 148$, legend in \figit{a}. Panels \figit{d} - \figit{f} Different system sizes at $T = 0.17$, legend in \figit{d}. Panels \figit{g} - \figit{i} Different system sizes at $T = 0.21$, legend in \figit{d}. }
	\label{fig:chimq}
\end{figure*}
$\chi_m(q)$ is plotted alongside its transversal and longitudinal components $\chi_{m\perp}(q)$ and $\chi_{m\parallel}(q)$ in Figure~\ref{fig:chimq} (recall that transversal and longitudinal is meant relative to the spontaneous magnetization in the system, cf. Section \ref{s:Models:Fluctuations}).

The first row compares results at different temperatures and a fixed system size $N = (256)^2$. It becomes clear that noticeable differences between $\chi_{m\perp}(q)$ and $\chi_{m\parallel}(q)$ only appear at low temperatures, since there is no macroscopic magnetization at high temperatures, voiding the distinction between longitudinal and transversal fluctuations. In the low temperature regime, however, finite-size effects allow for a well-defined finite magnetization. The distinction is valid there and leads to two different contributions to $\chi_m(q)$ that can be distinguished by their long wavelength behavior. It is to be expected that $\chi_{m\parallel}(q)$ approaches a finite value at small $q$ since it describes amplitude fluctuations in the local order parameter which come at a free energy cost. On the other hand, $\chi_{m\perp}(q)$, describing phase fluctuations in the local order parameter, i.e. its Nambu-Goldstone modes, is expected to diverge at low $q$ since the Hamiltonian is invariant under a uniform rotation of all spins. Consequently, the small $q$ behavior of $\chi_m(q)$ is dominated by the transversal contribution $\chi_{m\perp}(q)$ in the $q \to 0$ limit. Note also that for the system size $N=(256)^2$ under study, $\chi_{m\parallel}(q)$ becomes large and comparable to $\chi_{m\perp}(q)$ at the temperature $T = T_C(L) \approx 0.19$ that is related to a diverging correlation length.

To address the finite-size scaling of the $\chi_m(q)$ functions at and above the critical transition, we consider the lower two rows of Figure~\ref{fig:chimq}. We compare the susceptibilities at different system sizes for two fixed temperatures. One can immediately see from Panels~\figrefsub{fig:chimq}{\figit{d}-\figit{f}} that the scaling law \eqref{eq:ChiScaling} still holds at $T = 0.17$ and that the dominant long wavelength contribution to $\chi_{m}(q)$ is due to $\chi_{m\perp}(q)$. At $T = 0.21$ in Panels~\figrefsub{fig:chimq}{\figit{g}-\figit{i}}, however, scaling is lost. Small systems still show signatures of QLRO as $T_C(L) > 0.21$ for them, while for larger systems $T_C(L) < 0.21$ and the curves are no longer scale-free. These results are consistent with the analysis of the magnetization $\mean{m}$ and the correlation $C_m(r)$ discussed above.

The splitting of magnetization fluctuations into longitudinal and transversal parts is common practice for broken-symmetry systems, especially when the magnetization is conserved\cite{Perera2017}. Figure~\ref{fig:chimq} shows that this distinction is also valid for systems where the magnetization is neither strictly conserved nor is it an actual broken symmetry in the thermodynamic limit. The relevant feature is its decay over very large time scales. The  presence of strong finite-size effects discussed earlier makes the infinite system inaccessible to our simulation. As there cannot be fluctuations longitudinal to the magnetization in the thermodynamic limit due to the Mermin-Wagner theorem\cite{Mermin1966}, $\chi_{m\perp}(q)$ is the only relevant part of $\chi_{m}(q)$ when approaching the infinite systems. For that reason, we will mainly focus on fluctuations $C_{m\perp}(q,t)$ and spectra $S_{m\perp}(q,\omega)$ when discussing the dynamics in Section \ref{s:CoherentSpinACF}.

\subsubsection{The Exponent $\eta$}
\begin{figure}[htb]
	\includegraphics{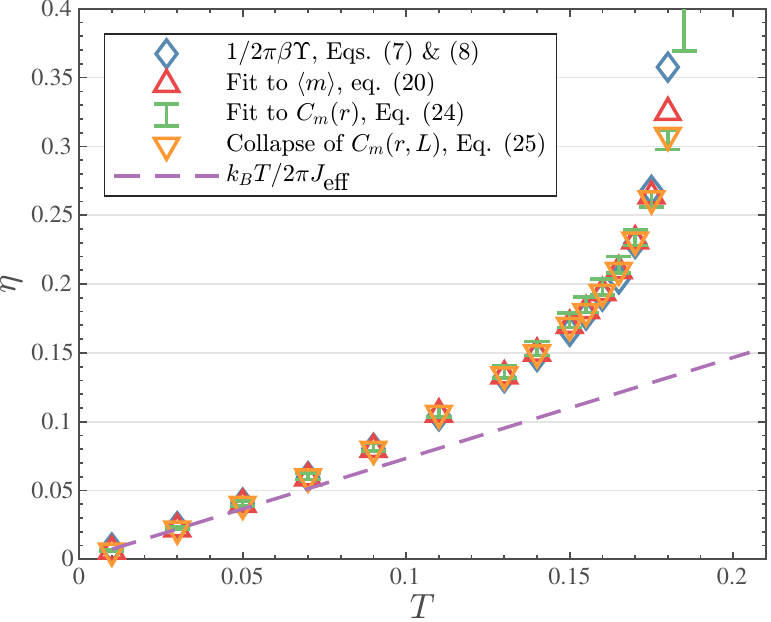}
	\caption{Exponent analysis for the MXY model. Critical exponent $\eta$ obtained from four different methods: the direct calculation of the helicity modulus $\Upsilon$ via \eqref{eq:Upsilon},\eqref{eq:H_x,I_X^2} and the relation $\eta = 1 /2\pi\beta\Upsilon$ (blue), from finite-size fits to the magnetization \eqref{eq:Magnetization} (red), power-law fits to $C_m(r)$ at $N=(256)^2$ \eqref{eq:SpinCorrAsymptotic} (green with error bars), and finite-size collapse fits to \eqref{eq:SpinCorrScaling} (orange). Symbols are roughly the size of errors bars. The purple dashed line describes a linear behavior at low temperatures, $\eta = \kT/(2\pi J\eff)$ see text.}
	\label{fig:SpinCorrExponents}
\end{figure}
To complete our discussion of the static properties, we compare different prescriptions for obtaining the exponent $\eta$. We saw above that one can obtain $\eta$ below $T_{\BKT}$ in four different ways, namely
\begin{enumerate}
	\item by directly calculating the helicity modulus $\Upsilon$ via \eqref{eq:Upsilon} and \eqref{eq:H_x,I_X^2} and $\eta = 1/2\pi\beta\Upsilon$,
	\item by fitting the finite-size magnetization $\mean{m}$ scaling described in \eqref{eq:M_SW},
	\item by fitting power laws to the correlation function $C_m$ as in \eqref{eq:SpinCorrScaling} at a fixed system size, say $N=(256)^2$, or
	\item by searching for an $\eta$ that optimizes data collapse according to Eq. \eqref{eq:SpinCorrScaling} in the correlation function when varying $N$.
\end{enumerate}
In Figure~\ref{fig:SpinCorrExponents}, these four methods are compared. The red error bars describe the quality of the fit to \eqref{eq:SpinCorrAsymptotic}, they do not account for statistical errors in the simulation data. Evidently, all methods mentioned arrive at similar exponents, especially the two finite-size fits \eqref{eq:M_SW} and \eqref{eq:SpinCorrScaling}. Within error bars, it is possible to fit a curve $\eta \sim a/(b + \sqrt{T-T_{\BKT}})$ to the data, in agreement with $\BKT$ theory\cite{Kosterlitz1974}.

One can also extract a critical temperature $T_{\BKT}$ from these fits by demanding $\eta = 1/4$. Using a spline interpolant, all three methods arrive at $T_{\BKT} = 0.173$. This result is consistent with the Bramwell-Holdsworth scaling analysis of the magnetization $\mean{m}$ in Section~\ref{s:Magnetization}, which also predicted $T_{\BKT} = 0.173$ at the largest system size.

At low temperatures, the renormalization of the spin wave stiffness $K$ due to vortex excitations is negligible. There, $K$ follows the prescription $K = J_{\textrm{eff}} / \kT$ with the effective spin-spin interaction strength $J_{\textrm{eff}}$. For an on-lattice model with nearest-neighbor spin-spin interaction strength $J$ and coordination number $Z$, $J_{\textrm{eff}} = J Z/4$.\cite{BissingerThesis2022,ChaikinLubensky2015} The orange line in Figure~\ref{fig:SpinCorrExponents} shows agreement with the simulation data up to $T \lesssim 0.06$, with the value $J_{\textrm{eff}} = 0.20(1)$. The standard XY model has a critical temperature of $T_{\BKT}/J_{\textrm{eff}} \approx 0.89$,\cite{Ueda2021} and indeed, we find $T_{\BKT}/J_{\textrm{eff}} = 0.85(9)$ with rather conservative error bars for the MXY model.

\subsection{Dynamics} 
\label{s:SimDyn}
In the previous section we found that the static properties of the MXY model are consistent with the universal properties of the standard XY model. Additionally, as we checked but did not show, the DXY model exhibits the identical static correlations. In order to understand the effect of mobility, we have to progress to dynamics, especially the properties of spin waves in the low temperature phase.

\subsubsection{Incoherent Spin Autocorrelation}
\label{s:IncoherentSpinACF}
We venture into the domain of time correlation functions by first studying $\Cminc(t) = \mean{\vs_i(0)\cdot\vs_i(t)}$, \eqref{eq:C^inc(t)}. Below $T_{\BKT}$, the correlation length $\xi$ is infinite, making the system size $L$ the relevant length scale for macroscopic fluctuations. By the dynamic scaling hypothesis,\cite{Hohenberg1977} one can assume universal behavior $\Cminc(t) = L^{-x}\hat{f}(t/L^z)$ with some exponents $x$ and $z$. Since $\Cminc(t) \to \mean{m}^2$ for $t\to\infty$, one must set $x=\eta$ with \eqref{eq:M_SW}.

A dynamic scaling calculation by Nelson and Fisher (NF)\cite{NelsonFisher1977} can be applied to $C_m^{\textrm{inc}}(t)$ to predict $C_m^{\textrm{inc}}(t) \sim t^{-\eta}$. The NF calculation is based on a Gaussian spin wave Hamiltonian for the spin angle variable. It is the basis for a phenomenological dynamical equation that can be solved analytically. The resulting correlation functions describe undamped spin waves at long wavelengths in an infinite system.

The NF law for $\Cminc(t)$ is supported by Lepri and Ruffo\cite{Lepri2001}, in whose theory it holds for short times $t/L \ll 1$, whereas finite-size effects lead to oscillatory behavior at $t/L \gg 1$. The oscillation is produced by long-wavelength propagating spin waves in a finite, periodically repeated box. Therefore, its period must scale linearly in $L$, which is tantamount to a dynamic critical exponent of $z=1$.

\begin{figure}[tb]
	\includegraphics[width=.95\columnwidth]{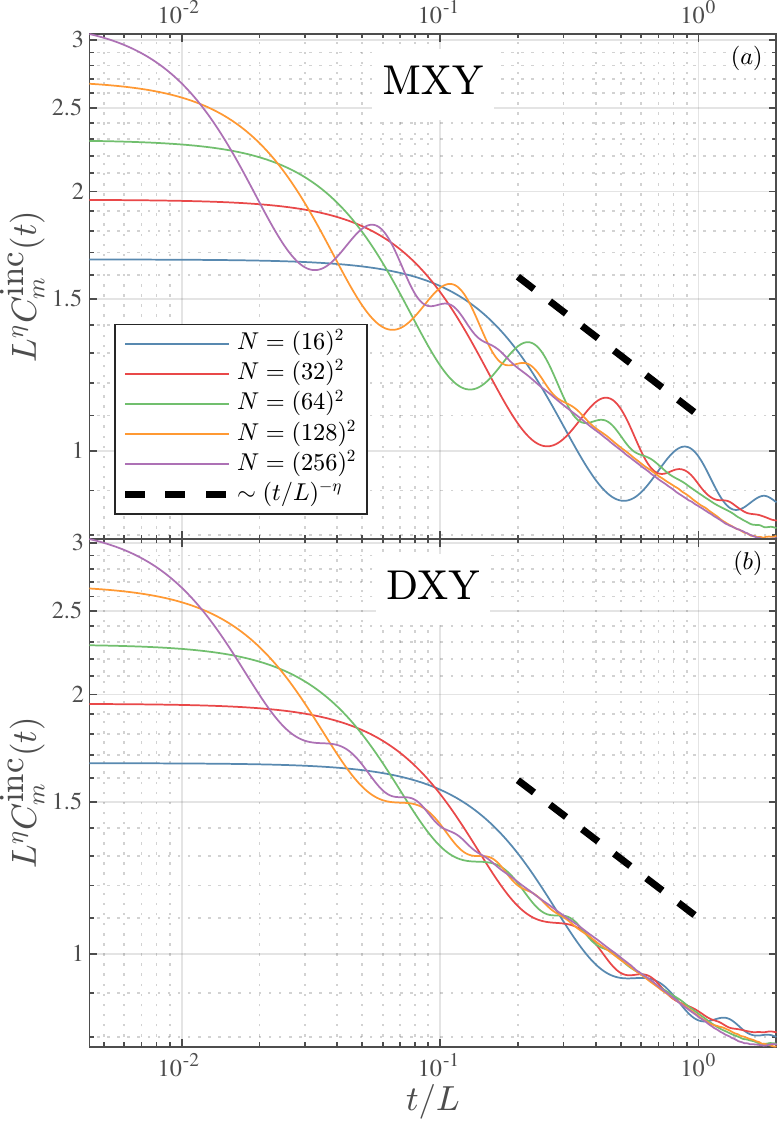}
	\caption{Finite-size plot of the short time behavior of $C_m^{\textrm{inc}}(t)$ at $T = 0.17$, where $\eta = 0.24$. Transient oscillations die down and the data collapses on a curve $\sim (t/L)^{-\eta}$, dashed line. Finite-size effects set in for longer times, see also Figure~\ref{fig:LepriRuffo}. }
	\label{fig:LepriRuffo_ShortTime}
\end{figure}
We analyze the short-time behavior of $C_m^{\textrm{inc}}(t)$ for the MXY and DXY model in Figure~\ref{fig:LepriRuffo_ShortTime}. The results do not fall on a single curve due to transient oscillations at short times. These arise from short time local interactions that are not described by the long wavelength low frequency spin wave theory of NF. There is data collapse after roughly the first three oscillations onto a single curve. This collapse cannot be observed for the smallest systems, since finite-size effects set in before the transients die down. The data agrees well with a universal curve $\sim t^{-\eta}$, plotted with a dashed line, where the exponent $\eta$ is the one obtained from the data collapse of Section \ref{s:SpinCorrRealSpace}. From this analysis, it appears that the NF scaling law also applies to both the MXY and DXY model at short times, and indeed curves as in Figure~\ref{fig:LepriRuffo_ShortTime} are valid throughout the low temperature regime with their corresponding values $\eta(T)$. One can observe NF scaling behavior up to $t/L \approx  1$ for both models.

\begin{figure}[tb]
	\includegraphics[width=\columnwidth]{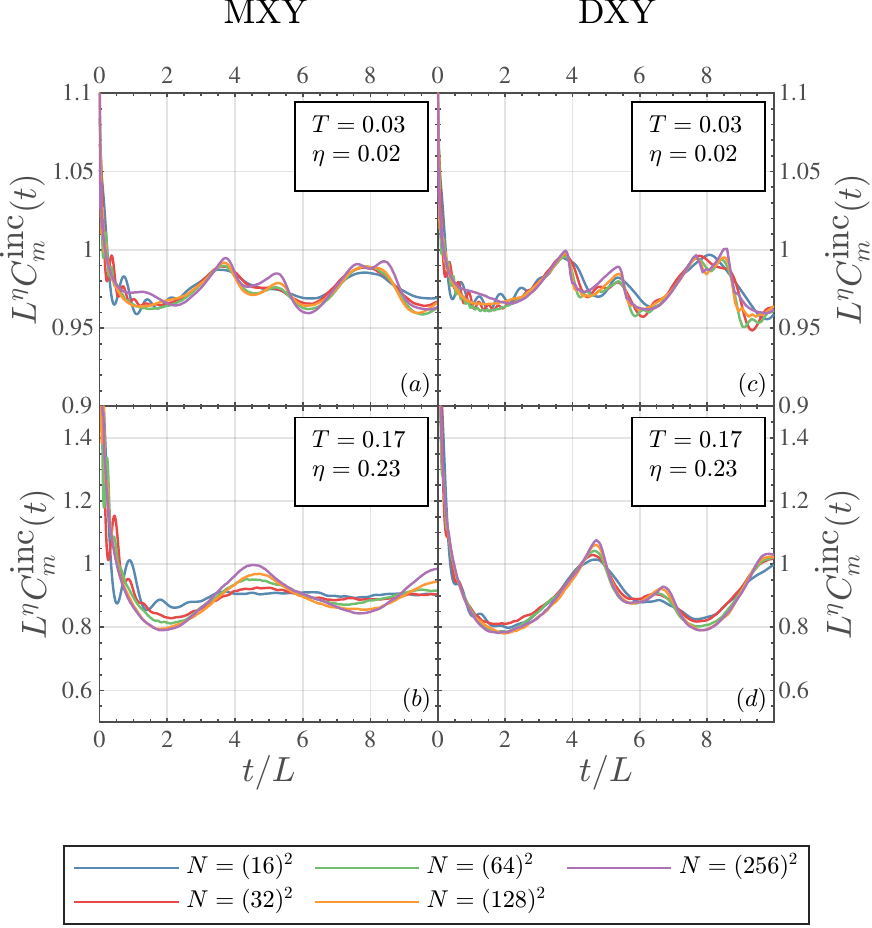}
	\caption{Spin wave behavior below (top) and around (bottom) the transition temperature for the MXY (left) and the DXY (right) model. }
	\label{fig:LepriRuffo}
\end{figure}
To address the finite-size oscillatory behavior at long times Lepri and Ruffo predicted due to their spin wave model\cite{Lepri2001}, we turn to Figure~\ref{fig:LepriRuffo}. The first column, Figures~\figrefsub{fig:LepriRuffo}{\figit{a},\figit{b}}, shows the results for the MXY model at different temperatures below the transition. One can see that there is some data collapse at the very low temperature $T = 0.03$, yet at higher temperatures, the oscillations do not seem to fit on a size-independent curve. The data collapse is stronger for the DXY model for both temperatures in Figures~\figrefsub{fig:LepriRuffo}{\figit{c},\figit{d}}.

A possible explanation for this difference is the different damping rates for spin waves in a mobile and a fixed-site model. The universal oscillations in the long-time behavior of $\Cminc(t)$ must be due to spin waves crossing the periodic boundary and leading to self-interactions of a spin with the spin wave it participated in earlier. Oscillations will be suppressed if the spin wave has partially dispersed before self-interaction. Thus, the ratio between the time scales of spin wave propagation and spin wave dispersion determines the strength of the finite-size oscillations. Both NF theory as well as the model discussed by Lepri-Ruffo neglect spin wave damping.

Physically, the spin waves in a mobile model will suffer stronger damping due to the additional translational degrees of freedom available for energy dissipation. Supporting this argumentation, the DXY model shows data collapse even in small systems. At very low temperatures, the dispersion time is much larger and we still find visible oscillatory behavior in Figure~\figrefsub{fig:LepriRuffo}{\figit{a}}). But in Figure~\figrefsub{fig:LepriRuffo}{\figit{b}}, the damping time for small systems is too short and the oscillatory behavior dies down quickly. We will show below that the spin wave propagation speed $c$ is size-independent (Figure~\figrefsub{fig:omegagamme_SpinWaveGammaExp}{\figit{a}}), leading to a propagation time scaling with $L$, while the damping scales with approximately $L^2$ (Figure~\figrefsub{fig:omegagamme_SpinWaveGammaExp}{\figit{b}}), so their ratio goes as $L\inv$. Indeed, the curves in Figure~\figrefsub{fig:LepriRuffo}{\figit{b}} appear to converge to a constant function with increasing $L$. Further investigation could clarify this point, especially the extent to which this curve can be considered universal.

\subsubsection{Coherent Spin Autocorrelation}
\label{s:CoherentSpinACF}
For a clearer picture of spin wave dynamics, we now focus on the collective correlation functions $C_{m\perp}(q,t)$, \eqref{eq:C_m(q,t)}, from which we can infer properties of the Goldstone modes (cf. discussion at the end of Section \ref{s:SpinCorrReciprocal}). We will also consider the power spectrum $S_{m\perp}(q,\om)$, \eqref{eq:S_m(q,om)}, and some related quantities.

Mathematically, $C_{m\perp}(q,t)$ is accessible to hydrodynamic approaches. From there, we expect to find the dynamics characteristic of a damped spin wave.\cite{Forster2018} In that case, a hydrodynamic (damped oscillator) fit of the form
\begin{equation}
	\frac{C_{m\perp}(q,t)}{\chimperpq} = e^{-\gamma t/2} \left(\cos(\om_1 t) + \frac{\gamma}{2\om_1}\sin(\om_1 t)\right)
	\label{eq:DampedOsciFit}
\end{equation}
should describe $C_{m\perp}(q,t)$, with the damping rate $\gamma$ and the oscillation frequency $\om_1^2 = \om_0^2 - \gamma^2/4$ depending on $q$ and $T$. We expect an exponential decay above the transition (imaginary $\om_1$). Fourier transformation yields
\begin{equation}
	\frac{S_{m\perp}(q,\omega)}{\chimperpq} = \frac{2\gamma \om_0^2}{(\om^2 - \om_0^2)^2 + \gamma^2\om^2}.
	\label{eq:DampedOsciFit_S}
\end{equation}
The solutions \eqref{eq:DampedOsciFit} and \eqref{eq:DampedOsciFit_S} correspond to a Markovian approximation of the memory kernel appearing in e.g. the Zwanzig-Mori\cite{Forster2018,Zwanzig2001} treatment of the transversal magnetization fluctuations. Errors in this approximation are due to assuming frequency-independence for the damping rate $\gamma(\omega) = \gamma$ in \eqref{eq:DampedOsciFit_S}, corrections to this approximation typically start with adding details here. Fits were performed using standard MATLAB functions. For the Fourier transform involved in \eqref{eq:DampedOsciFit_S}, we use \eqref{eq:FourierTrafoWithConvolution}, which requires an additional time scale $\tau$. To determine $\tau$, we first obtained $\gamma$ from a fit to the time domain function \eqref{eq:DampedOsciFit}. Noise becomes dominant when $e^{-\gamma t/2} = \nruns^{-1/2}$, which leads to $\tau = \ln(\nruns)/\gamma$ for the calculation of the Fourier transform to obtain $S_{m\perp}(q,\omega)$.

We can characterize $S_{m\perp}(q,\omega)$ by the height and position of its maximum
\begin{equation}
	\begin{aligned}
		\omega_{\max}(q) &=  \arg\max_{\omega} S_{m\perp}(q,\omega),\\
		S_{m\perp}^{\max}(q) &= \max_{\omega} S_{m\perp}(q,\omega)
	\end{aligned}
\end{equation} 
The fit \eqref{eq:DampedOsciFit_S} predicts $S_{m\perp}^{\max} = 2\chi_{m\perp}/\gamma$ provided that $\omega_1^2 \approx \omega_0^2$. The peak position is at $\omega_{\max}^2 = \omega_0^2 - \gamma^2/2 = 2\omega_1^2 - \omega_0^2$ and shifts to zero if $\omega_{\max}^2 < 0$. Damping rates are often extracted from the full width at half maximum (FWHM) or the half width at half maximum (HWHM) of the peak. We will study the damping rates predicted by $S_{m\perp}^{\max}$ and the FWHM below.

\begin{figure*}[htb]
	\includegraphics[width=\linewidth]{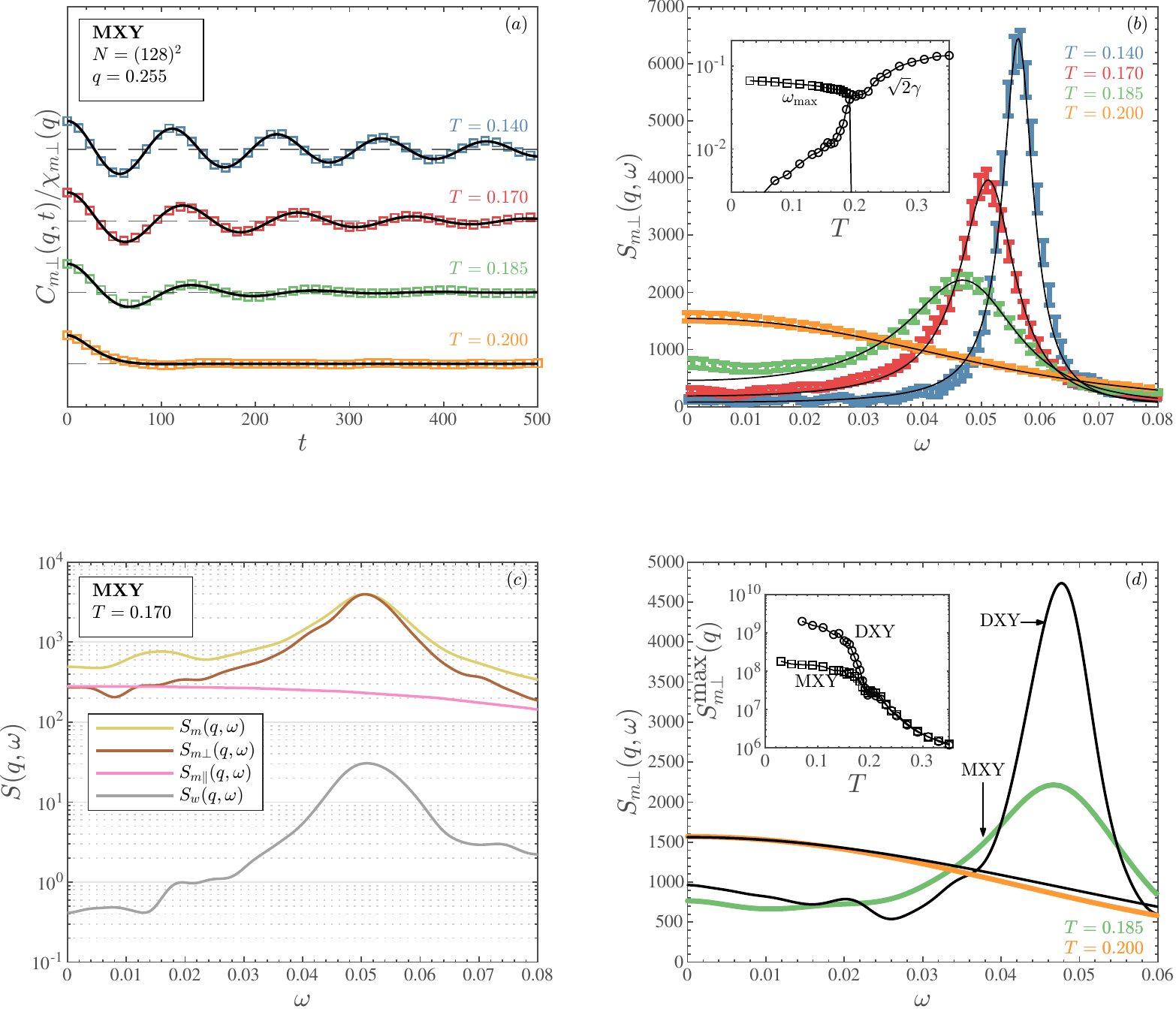}
	\caption{Properties of coherent time correlation functions at system size $N = (128)^2$ and wavevector $q = 0.255$. \figit{a} Spin correlation function $C_{m\perp}(q,t)$ in the time domain at different temperatures. Symbols are simulation data, solid lines are fits to \eqref{eq:DampedOsciFit}. Plots are shifted and normalized for better visibility. Errors are smaller than the symbols. \figit{b} The corresponding power spectrum $S_{m\perp}(q,\omega)$. Error bars for simulation data, solid line for a fit to \eqref{eq:DampedOsciFit_S}. The inset compares the peak position $\omega_{\max}$ to the damping rate $\gamma$ over a wider temperature range. $\omega_{\max}$ jumps to zero at $T_C = 0.19$. \figit{c} Power spectra of the three different spin correlations and the spin momentum correlation. \figit{d} Comparison of $S_{m\perp}(q,\omega)$ between the MXY and the DXY model. The inset compares the peak height $S_{m\perp}^{\max}(q)$ over a wider temperature range.}
	\label{fig:SpinTCF_DataAndFit}
\end{figure*}
Figure~\ref{fig:SpinTCF_DataAndFit} gives an impression of the behavior of $C_{m\perp}(q,t)$ and $S_{m\perp}(q,\omega)$. Figures~\figrefsub{fig:SpinTCF_DataAndFit}{\figit{a},\figit{b}} present simulation data (symbols) alongside fits to \eqref{eq:DampedOsciFit} and \eqref{eq:DampedOsciFit_S} (solid lines), respectively. It is hard to spot shortcomings of the hydrodynamic fit in the time domain. In the frequency domain, the fit is capable of accurately describing the main spin wave peak. At lower frequencies, there are some disagreements, especially for the temperature $T= 0.185$ that lies within the transition region, which shows the emergence of an additional central peak that is not covered in the Markovian damped oscillator model \eqref{eq:DampedOsciFit_S}. For temperatures below the transition, there are slight deviations from the monotonous behavior at $\omega < \omega_{\max}$ of \eqref{eq:DampedOsciFit_S}. Whether these are due to multi-spin scattering as Evertz and Landau\cite{Evertz1996} proposed in their Monte Carlo analysis cannot be decided within the error margins of our data and we will not discuss this matter. As for the high frequency behavior, we will come back to that shortly.

The inset in Figure~\figrefsub{fig:SpinTCF_DataAndFit}{\figit{b}} allows for a new perspective on what happens during the phase transition. The characteristic frequency of the system, described here by $\omega_{\max}$, the position of the maximum in $S_{m\perp}(q,\om)$, changes only minimally in the temperature range below the transition, while the damping rate $\gamma$, starting out from very small values at small $T$, increases strongly with temperature. There is a temperature where $\sqrt{2}\gamma$ intersects $\omega_{\max}$, which is immediately followed by a strong decline in $\omega_{\max}$. This is the transition to the overdamped regime around $T = T_C(N = 128) = 0.195$. In the high temperature phase, propagating spin waves cease to exist and spin diffusion sets in.

Figure~\figrefsub{fig:SpinTCF_DataAndFit}{\figit{d}} is a cut from Figure~\figrefsub{fig:SpinTCF_DataAndFit}{\figit{b}} that focuses on the curves at $T = 0.17$ and $T= 0.20$ and compares the spin wave dynamics in the MXY model to that in the DXY model. Evidently, the difference is hardly noticeable in the high temperature phase, while the spin wave peak of the low temperature phase is significantly more pronounced in the DXY model. The position $\omega_{\max}$ of the peak, however, does not shift. The inset of Figure~\figrefsub{fig:SpinTCF_DataAndFit}{\figit{d}} follows this trend over a wider range of temperatures, and indeed, $S_{m\perp}^{\max}$ (which in the damped oscillator model is just inversely proportional to $\gamma$) agrees for the two models above the transition while differing by roughly an order of magnitude below it. 

We may also analyze the spin waves by means of other correlation functions, which we study in Figure~\figrefsub{fig:SpinTCF_DataAndFit}{\figit{c}}. As before in the analysis of Figure~\ref{fig:chimq}, we observe that there are strong differences in $S_{m\perp}(q,\omega)$ and $S_{m\parallel}(q,\omega)$ in the spin-wave regime. While the former shows the spin wave peak, the latter has a Lorentzian peak. They do agree at $\omega = 0$. The low $q$ behavior of $S_{m}(q,\omega)$ is mainly determined by $S_{m\perp}(q,\omega)$, especially around the spin wave frequency $\omega_{\max}$. Interestingly, the spin momentum fluctuations also have a peak in their power spectrum $S_{w}(q,\omega)$ at $\omega_{\max}$ with very similar width and relative height. This is not trivial, since there is no continuity equation, $\dot{m}_{\perp,\qq} \neq -iq\wq$. Still, it is only reasonable to assume some form of reactive coupling between $m_{\perp,\qq}$ and $\wq$, manifest in them participating jointly in the spin wave.

\begin{figure}[tb]
	\includegraphics{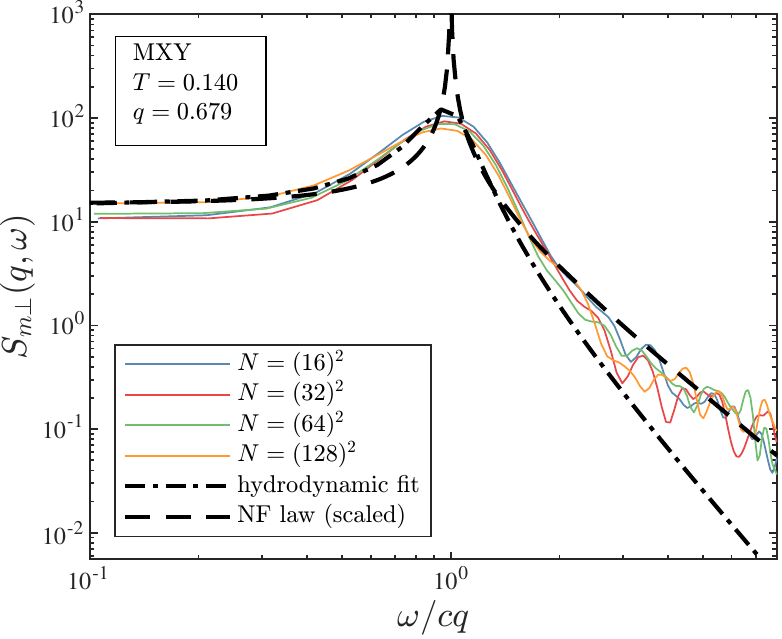}
	\caption{Peak structure and high-frequency behavior of the MXY model at different system sizes. The dash-dotted line is the hydrodynamic fit to the $N=(128)^2$ data, the dashed line is the theoretical result by Nelson and Fisher\cite{NelsonFisher1977} (arbitrarily scaled to agree at $\omega = 0$). Data has been artificially smoothened for a better visibility of high-frequency trends.}
	\label{fig:NelsonFisher}
\end{figure}
In our discussion of the incoherent correlations $\Cminc(t)$ in Figures~\ref{fig:LepriRuffo_ShortTime} and \ref{fig:LepriRuffo}, we found that the theory for dynamic correlations by Nelson and Fisher\cite{NelsonFisher1977} only applies for short-time properties. Similarly, we want to take a closer look at the high $\omega$ behavior of the spin fluctuation power spectra. By the Nelson-Fisher theory, the function $cqS_{m}(q,\omega)/\chi_{m}(q)$ is universal (see also Evertz and Landau\cite{Evertz1996}). This should especially apply to $S_{m\perp}(q,\omega)$ in the finite system case. To study that, we turn to Figure~\ref{fig:NelsonFisher}. Our observations are threefold: for once, we observe that the data agree for different system size within statistical error (error bars not shown). Secondly, the peak shape is better described by our hydrodynamic fit than the shape proposed by Nelson and Fisher. On the other hand, the high-frequency behavior is captured more accurately by Nelson-Fisher theory than by a \eqref{eq:DampedOsciFit_S}. This shows again that in a time window $t_1 < t < t_2$, the correlation decays as $t^{-\eta}$, where $t_1$ is a characteristic time for transient effects while $t_2$ is connected to the crossover to the hydrodynamic regime.

\begin{figure}
	\centering
	\includegraphics{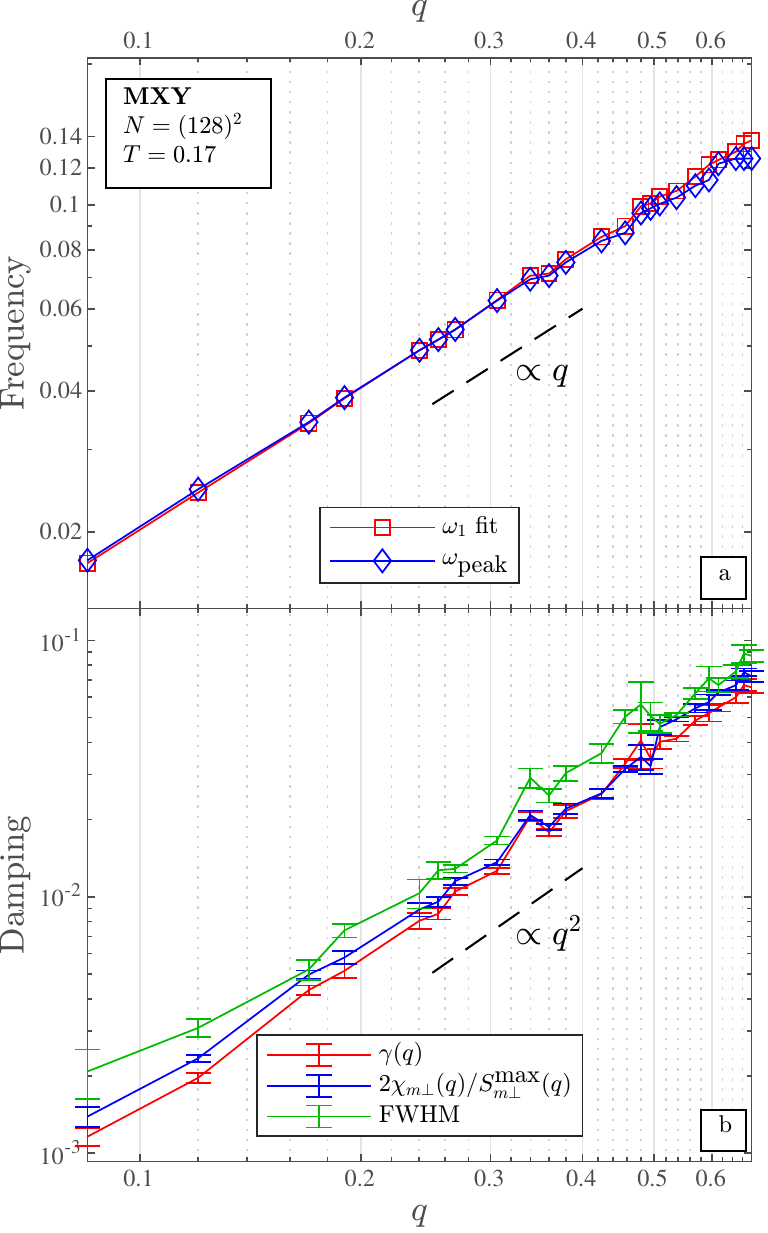}
	\caption{Comparison of strategies to obtain frequency and damping coefficients. }
	\label{fig:omegagamme_fitcompare}
\end{figure}
We now turn to the coefficients $\omega_1$ and $\gamma$ or their frequency domain counterparts $\omega_{\max}$ and $S_{m\perp}^{\max}$ and the FWHM in Figure~\ref{fig:omegagamme_fitcompare}. For a temperature within the spin wave region, $\omega_1$ agrees very well with $\omega_{\max}$. The agreement between $\gamma$, $2\chi_{m\perp}/S_{m\perp}^{\max}$ and the FWHM is not as strong, but given the fit uncertainty in $\gamma$ and the error bars on the peak values $S_{m\perp}^{\max}$ in Figure~\figrefsub{fig:SpinTCF_DataAndFit}{\figit{b}} as well as the comparable errors in $\chi_{m\perp}$ (which we have not explicitly given), the discrepancy can be considered to lie within error bars. The FWHM is only expected to be proportional to the damping rate, therefore exact agreement is also not to be expected. This allows us to conclude that we can use the time domain as well as the frequency domain as viable tools for analyzing the spin wave damping. From a computational point of view, this is especially useful, since the time domain fit does not require one to sample over the entire time of the order $\gamma\inv$, which would be required to get a good resolution of the peak in the power spectrum.

We can also use the $q$-dependence of $\omega_1$ to study the dynamical critical exponent $z$ of the model. For that, we consider the finite-size scaling form\cite{Hohenberg1977,NelsonFisher1977,Evertz1996}
\begin{equation}
	\omega_{1}(q,L) = q^z \hat{\omega}_{1}(qL)
\end{equation}
The universal function $\hat{\omega}_{1}(qL)$ will still depend on temperature. It is quite clear from Figure~\figrefsub{fig:omegagamme_fitcompare}{\figit{a}} that $\omega_1 = cq$ describes acoustic excitations, which is very natural for a propagating mode like the spin waves. Therefore, $z=1$, which is in agreement with theoretical\cite{NelsonFisher1977} and simulation\cite{Evertz1996} findings of the easy-plane Heisenberg magnets.

\begin{figure}
	\centering
	\includegraphics{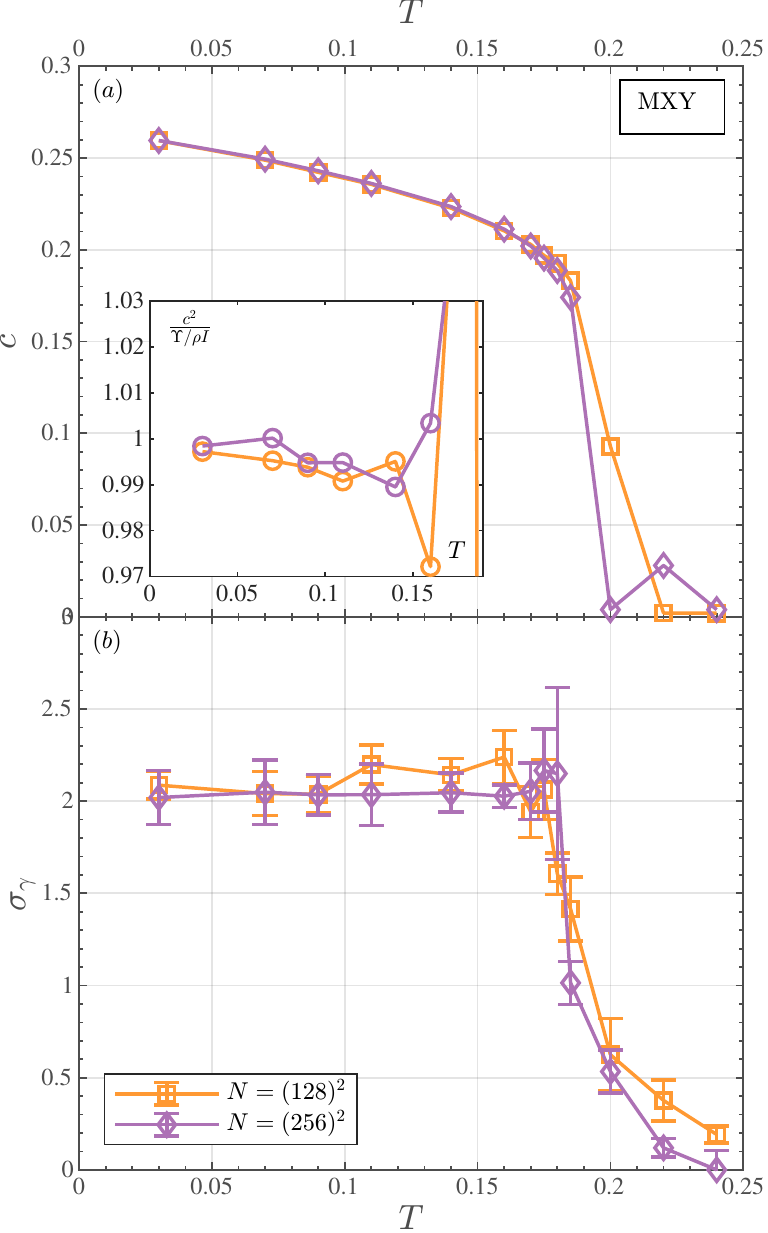}
	\caption{Study of the power-law behavior of $\omega_1$ and $\gamma$ from fits to $C_{m\perp}(q,t)$ at system size $N=(128)^2$ and $N=(256)^2$. \figit{a} Spin wave propagation speed $c$ computed by $\omega_1 = cq$. The inset shows the ratio $c^2/(\Upsilon / (\rho I))$. \figit{b} Power law exponent for $\gamma = aq^{\sigma_{\gamma}}$.}
	\label{fig:omegagamme_SpinWaveGammaExp}
\end{figure}
Figure~\ref{fig:omegagamme_fitcompare} indicates that both $\omega_1$ and $\gamma$ follow power laws in $q$ at $T= 0.17$. Indeed, we observe similar behavior for all $T < T_{\BKT}$. Given the dispersion relation for $\omega_1$, this is not surprising, yet it is an interesting finding concerning $\gamma$, since this implies that the magnetization becomes quasi-conserved in the limit $q \to 0$, i.e. $L \to \infty$. Note that this statement is compatible with the Mermin-Wagner theorem's demand $\mean{m} = 0$. A quasi-conserved magnetization is in disagreement with the theory of Mertens \textit{et al.}\cite{Mertens1989} Differences could be due to the mobility of the particles or differences between rotator XY models and easy-plane Heisenberg magnets. Also, Mertens \textit{et al.}\cite{Mertens1989} did not focus their investigation on the low $q$ behavior of the spin wave damping, thus their data leaves room for interpretation concerning the low $q$ trend.

Finally, Figure~\ref{fig:omegagamme_SpinWaveGammaExp} studies the power-law behavior of $\omega_1$ and $\gamma$. Figure~\figrefsub{fig:omegagamme_SpinWaveGammaExp}{\figit{a}} shows the spin wave propagation speed $c$, computed from $\omega_1 = cq$ for low $q$, obtained for different temperatures at the two largest system sizes. Clearly, the result is only $N$-dependent in the finite-size transition region $T^* < T < T_C$. We see convergence to a jump in $c$ around $T_{\BKT}$, which is consistent with the jump in spin wave stiffness predicted by BKT theory\cite{Berezinskii1971destruction,Kosterlitz1974}. The value $c = 0$ above the transition is due to the appearance of the central peak in $S_{m\perp}(q,\omega)$. Also, within a Zwanzig-Mori projection operator approach\cite{Zwanzig2001}, one can derive $c^2 = \Upsilon / (\rho I)$ for an MXY model under the assumption of a homogeneous density.\cite{BissingerThesis2022} The inset of \figrefsub{fig:omegagamme_SpinWaveGammaExp}{\figit{a}} confirms this result, and the helicity modulus $\Upsilon$ calculated from \eqref{eq:Upsilon} and \eqref{eq:H_x,I_X^2} indeed shows the correct behavior.

If we assume a simple power law $\gamma = a q^{\sigma_\gamma}$, we can fit the curves $\gamma(q)$ for each temperature and obtain results collected in Figure~\figrefsub{fig:omegagamme_SpinWaveGammaExp}{\figit{b}}. The data is noisier and error bars are larger, since we are now using a curve fit on the result of a curve fit. However, there is a clear trend to values of $\sigma_{\gamma}$ around $\sigma_{\gamma} = 2$ in the spin wave region $T < T_{\BKT}$ to vanishing values above the transition, where a fit $\gamma = b + a q^{\sigma_{\gamma}}$ should be more suitable. While there is a noticeable dependence of $\sigma_{\gamma}$ on the system size in Figure~\figrefsub{fig:omegagamme_SpinWaveGammaExp}{\figit{b}}, the data mostly agree to within error bars. We also found noise to decrease with system size, which is why we only show the largest two system sizes under study in Figure~\ref{fig:omegagamme_SpinWaveGammaExp}. 

In this context, one could expect the exponent $\sigma_{\gamma} = \sigma_{\gamma}(\eta)$ to depend on the exponent $\eta$ linked to the coupling constant in the effective Hamiltonian of the system,\cite{Kosterlitz1974} and indeed the data becomes more ambiguous close to the transition temperature, where spin wave renormalization of $\eta$ becomes relevant. A more careful investigation of a larger set of data would be required for a quantitative analysis.

\section{Conclusions}
\label{s:Conclusion}
In this paper, we provided evidence that a mobile XY model with purely repulsive interaction undergoes a BKT phase transition very much akin to the static XY model. The finite-size scaling analysis of Bramwell and Holdsworth\cite{Bramwell1993,Bramwell1994magnetization} was carried out and leads to similar results with the finite-size critical exponents $\beta$ in agreement with the BKT scenario. The transition temperature from this analysis is $T_{\BKT} = 0.17(1)$ at a density $\rho = 2.99$. It is evident that the phase transition is not affected by the positional ordering of the system, since crystalline structures break down at much lower temperatures. The fluctuation of the absolute magnetization $\chi_{m}$ is also in agreement with the BKT transition, including the correct finite-size scaling of its maximum value, as well as the universal properties of the magnetization probability distribution function $P(m)$. Further finite-size scaling analysis was performed on the static correlations $C_m(r)$ and $\chi_m(q)$, providing estimates for the exponent $\eta$ and showing failure of data collapse above $T \gtrsim 0.19$. Gathering these different approaches, we find a jump value of the critical exponent $\eta_\BKT$ close to $\eta_\BKT = 1/4$ at $T_{\BKT}$. We conclude from this collection of static properties that the purely repulsive MXY model indeed undergoes a BKT transition.

Studying the dynamics of the model, we found the Nelson-Fisher law\cite{NelsonFisher1977} for the incoherent spin autocorrelation function $\Cminc(t)$ confirmed for short times. The long-time finite-size effects predicted by Lepri and Ruffo\cite{Lepri2001}, however, took different forms for the mobile model. We assume that this is due to additional diffusive transport processes being relevant at small system sizes. To our knowledge, this is the first simulation confirmation of the results of Lepri and Ruffo\cite{Lepri2001}, which hold for the DXY model.

Finally, we verified the existence of damped spin waves in the coherent transversal spin autocorrelation function $C_{m\perp}(q,t)$ as well as its power spectrum $S_{m\perp}(q,\omega)$ for $T < T_C$. The hydrodynamic law enters an overdamped regime above the finite-size transition temperature $T_C$, while it cannot cover the appearance of a central peak in the region $T^* < T < T_C$. In the spin wave phase, the hydrodynamic fit fails to describe the high frequency behavior of $S_{m\perp}(q,\omega)$, which is in better agreement with the theoretical results by Nelson and Fisher\cite{NelsonFisher1977}.

We found that a coefficient fit in the time domain gives similar results as an analysis of the shape and position of the peak in the frequency domain. Comparing the MXY and the DXY model, we found that the characteristic frequency $\omega_{1}(q)$ agrees for both models, whereas the damping rate $\gamma(q)$ is much larger for $T < T_{\BKT}$ in the DXY model, while it agrees for both models in the disordered phase.

Concerning the dependence of the fit parameters $\omega_1(q)$ and $\gamma(q)$ on $q$, we find that $\omega_1(q)$ describes an acoustic mode, $\omega_1(q) = cq$ at low $q$. With that, we confirm agreement of the dynamical critical exponent $z=1$ with that of the standard XY model. Also, the spin wave frequency $c$ shows the expected jump discontinuity at the transition. The relationship $c^2 = \Upsilon/\rho I$ found in theory\cite{BissingerThesis2022} is verified to good agreement by the simulation data. The damping rate fit parameter $\gamma(q)$ can also be described by a power law $\gamma(q) = aq^{\sigma_{\gamma}}$ at low $q$ in the spin wave phase. The choice $\sigma_{\gamma} = 2$ agrees with the data, which implies that spin waves in the transverse component of the magnetization become quasi-conserved quantities. We checked that this result agrees with the standard XY model.

On a methodological level, we used a split of the magnetization fluctuations into parts parallel and perpendicular to the instantaneous magnetization $\mm$, $m_{\parallel,\qq}$ and $m_{\perp,\qq}$. We showed that in both the static susceptibility $\chi_{m}(q)$ and the dynamic autocorrelation $C_{m}(q,t)$, the transversal fluctuations are the ones relevant to the critical properties of the system. 

\section*{Acknowledgement}
We are thankful to Mathias Höfler for discussions and his contributions to the simulation environment. Furthermore, we wish to acknowledge funding by the DFG within SFB 1432 (ID 425217212).

\appendix
\section{Obtaining the Helicity Modulus from a Twist Field}
\label{c:AppendixHelicity}
The expressions \eqref{eq:Upsilon} and \eqref{eq:H_x,I_X^2} can be derived analytically. The derivation here is in close analogy to a well-known proof (for a didactic explanation, see Sandvik\cite{Sandvik2010computational}, whose structure we follow).

For static systems, expressions for the helicity modulus are obtained by introducing a twist into the system such that $\nab \te = \phi \ee_x$ minimizes the free energy, \cite{ChaikinLubensky2015,Sandvik2010computational} and obtain $\Upsilon$ by\cite{BissingerThesis2022}
\begin{equation}
	\Upsilon = \frac{1}{A} \parderi{^2 F}{\phi^2}
	\label{eq:GEN:UpsilonDef}
\end{equation}
with $A = L_x L_y$. It can be related to the spin wave coupling $K$ via $K = \beta \Upsilon$.\cite{BissingerThesis2022}

Instead of actually enforcing a twist on the system, an analytic expression for $\Upsilon$ can be calculated by introducing a twist field $\Phi(\rr)$ into the spin-spin interaction and then taking a small $\Phi$ expansion to arrive at an expression for $\Upsilon$ that is based on simple static averages.\cite{Sandvik2010computational} We can change the spin interaction term in the Hamiltonian \eqref{eq:MXY} to
\begin{equation}
	H(\Phi)
	= \sum_{i\neq j} J(r_{ij})\cos(\te_{ij} - (\Phi(\rr_i) - \Phi(\rr_j))).
	\label{eq:GEN:MXY_Hamiltonian}
\end{equation}
The twist field $\Phi(\rr)$ distorts the nearest-neighbor interaction (with $\rr = (x,y)\transp$). We shall choose $\Phi(\rr) = \phi x$, which would enforce the $\te_i = \phi x_i$ at $T = 0$ ($x_{i} = \rr_{i} \cdot \vec{e}_x$ and similarly $x_{ij} = \rr_{ij}\cdot\vec{e}_x$), and leads to the interaction part of the Hamiltonian containing $\cos(\te_{ij} - x_{ij} \phi)$. By the trigonometric identity $\cos(a-b)=\cos(a)\cos(b) + \sin(a)\sin(b)$ and an expansion to lowest order with $\cos(x) = 1 - x^2/2$ and $\sin(x) = x$, we arrive at an overall energy
\begin{equation}
	H(\phi) = H_0 + \frac{1}{2} H_x \phi^2 - I_x \phi.
\end{equation}
with
\begin{equation}
	\begin{aligned}
		H_x &= \frac{1}{2}\sum_{i \neq j } J(r_{ij})\cos(\te_{ij}) x_{ij}^2,\\
		I_x &= \frac{1}{2}\sum_{i \neq j } J(r_{ij})\sin(\te_{ij}) x_{ij},
	\end{aligned}
\end{equation}
and $H_0 = H(\phi = 0)$ is just the standard MXY Hamiltonian \eqref{eq:MXY} without a twist field. Expanding the partition function to lowest order in $\phi$ leads to a free energy\cite{BissingerThesis2022,Sandvik2010computational}
\begin{equation}
	F = -\frac{1}{\beta} \ln Z = F_0 + \frac{1}{2}\phi^2 \left(\mean{H_x} - \beta \mean{I_x^2} \right) + {\cal O}(\phi^4)
\end{equation}
with $F_0$ arising due to the Hamiltonian without external field. This is the lowest order response of the free energy to the twist field $\Phi(\rr)$. The helicity modulus $\Upsilon$ can be determined by \eqref{eq:GEN:UpsilonDef}. The direction $x$ of the twist field $\Phi$ was a wanton choice, one could just as well have chosen $y$ -- the important matter is that it is a symmetry direction of the system with respect to its periodic boundaries. Therefore, analogous quantities $H_y$ and $I_y$ can be defined and the helicity becomes
\begin{equation}
	\Upsilon
	= \frac{1}{A} \parderi{^2 F}{\phi^2}
	= \frac{1}{2A}\left(\mean{H_x + H_y} - \beta \mean{I_x^2 + I_y^2} \right).
	\label{eq:GEN:UpsilonFinal}
\end{equation}

\bibliography{./bibfile}

\end{document}